\documentclass[journal,draftcls,onecolumn,12pt,twoside]{IEEEtranTCOM}
\renewcommand{\baselinestretch}{1.72}

\normalsize
\IEEEoverridecommandlockouts
\usepackage{graphicx}
\usepackage{textcomp}
\def\BibTeX{{\rm B\kern-.05em{\sc i\kern-.025em b}\kern-.08em
		T\kern-.1667em\lower.7ex\hbox{E}\kern-.125emX}}
\usepackage{ulem}
\usepackage{amsmath}
\usepackage{amsthm,amssymb,amsmath,bm}
\usepackage{subfigure}
\usepackage{amsfonts}
\usepackage{epsfig}
\usepackage{amssymb}
\usepackage{amsmath}
\usepackage[table,xcdraw]{xcolor}
\usepackage{cite}
\usepackage[utf8x]{inputenc}
\usepackage{color,soul}
\usepackage{subfigure}
\usepackage{multirow}
\usepackage{rotating}
\usepackage{graphicx}
\usepackage{tabularx}
\usepackage{array}
\usepackage{color,soul}
\usepackage{bm}
\usepackage{graphicx,dblfloatfix}
\usepackage{blindtext}

\newtheorem{remark}{Remark}
\newtheorem{Defi}{Definition}
\usepackage{subfigure}
\usepackage{moreverb}
\usepackage{epsfig}
\usepackage{amsmath,amssymb,amsthm,mathrsfs,amsfonts,dsfont}
\usepackage{adjustbox,lipsum}
\usepackage{amsfonts}
\usepackage{epsfig}
\usepackage{amssymb}
\usepackage{amsmath}
\usepackage{amsthm}
\usepackage{subfigure}
\usepackage{multirow}
\usepackage{rotating}
\usepackage{graphicx}
\usepackage{tabularx}
\usepackage{array}
\usepackage{anyfontsize}
\usepackage{color,soul}
\usepackage{graphicx,dblfloatfix}
\usepackage{epstopdf}
\usepackage{blindtext}
\usepackage{amsmath}
\usepackage{amsthm,amssymb,amsmath,bm}
\usepackage{subfigure}
\usepackage{amsfonts}
\usepackage{epsfig}
\usepackage{amssymb}
\usepackage{amsfonts}
\usepackage{amsmath}
\usepackage{cite}
\usepackage{graphicx}
\usepackage{fancyhdr}
\usepackage{subfigure}
\usepackage[noblocks]{authblk}
\usepackage[subfigure]{tocloft}
\usepackage[font={small}]{caption}
\usepackage{subfigure}
\usepackage{tabularx}
\usepackage{tcolorbox}
\usepackage{cite}
\usepackage[linesnumbered,ruled,vlined]{algorithm2e}
\SetKwInput{KwInput}{Input}
\SetKwInput{KwOutput}{Output}
\usepackage{amsthm,amssymb,amsmath,bm}
\usepackage{graphicx}
\usepackage{fancyhdr}
\usepackage{subfigure}
\usepackage[subfigure]{tocloft}
\usepackage[font={small}]{caption}
\usepackage{subfigure}
\usepackage{tabularx}
\usepackage{cite}
\usepackage{hyperref}
\begin{document}
	\title{
		A Unified Framework for Joint  Energy and AoI Optimization via Deep Reinforcement Learning for 
		 NOMA   MEC-based Networks
		}
		
	\author{Abolfazl Zakeri, \textit{Student Member, IEEE},~Mohammad Parvini,~Mohammad Reza Javan,~\textit{Senior Member, IEEE},~Nader Mokari,~\textit{Senior Member, IEEE}, and  Eduard A Jorswieck, \textit{Fellow, IEEE}
		\thanks{A. Zakeri, M. Parvini,~and N. Mokari are with the Department of Electrical and Computer Engineering,~Tarbiat Modares University,~Tehran,~Iran, (e-mails: \{abolfazl.zakeri, m.parvini, nader.mokari\}modares.ac.ir). M.~R.~Javan is with the Department of Electrical and Computer Engineering, Shahrood University of Technology, Iran, (e-mail: javan@shahroodut.ac.ir).~Eduard A. Jorswieck is with Institute for Communications Technology, TU
			Braunschweig, Germany, Email: jorswieck@ifn.ing.tu-bs.de.
	}}
	\maketitle
		\vspace{-2cm}
	\begin{abstract}
		In this paper, we design a novel scheduling and resource allocation algorithm for a smart mobile edge computing (MEC) assisted radio access network.
		 Different from previous energy efficiency (EE)-based or average age of information (AAoI)-based network designs,
		we propose a unified  metric for 
		  simultaneously optimizing ESE and AAoI of the network. To further improve the system capacity, non-orthogonal multiple access (NOMA) is proposed as a candidate multiple access scheme for future  cellular networks. Our main aim  is maximizing the long-term objective function under AoI, NOMA, and resource capacity constraints using stochastic optimization. To overcome the complexities and unknown dynamics of the network parameters (e.g., wireless channel and interference), we apply the concept of reinforcement learning and implement a deep Q-network (DQN). Simulation results illustrate the effectiveness of the proposed framework and analyze different parameters’ impact on the network performance. Based on the results, our proposed reward function converges fast with negligible loss value.  Also, they illustrate our work outperforms the existing state of the art baselines up to  64\% in objective function and 51\% in AAoI, which are stated as examples.			
	\end{abstract}
	\section{introduction}
Fifth-generation (5G) wireless networks go live now and sixth-generation (6G) future networks are under research to provide ever-new services and applications besides personal assistant devices such as the internet of everything (IoE) applications in an efficient and intelligent manner \cite{Zakeri, Benis_6G}. 6G will undergo an unprecedented transformation that will make it significantly different from the previous generations of the wireless cellular networks. Accordingly, 6G needs new emerging performance metrics and zero-touch resource scheduling methods with a seminal effect on the network performance and  future marketing. In this regard, the age of information (AoI) \cite{AoI_IoT_Mag} and machine learning (ML)-based algorithms, especially reinforcement learning (RL) are  proposed  as a new network freshness  metric and network-level intelligence, respectively. 
	They are pivotal for the resource management  and orchestration to enable services with diverse quality of service (QoS) requirements using the limited and expensive network resources \cite{Infocom_1}.
 Notably,   recent attentions are in the area of power-saving and efficient spectrum utilization to improve the battery life of the machine-type-communication era \cite{Energy_IoT}.
\\\indent 
	Nevertheless, designing high-capacity,  efficient, and fast convergence ML  algorithms for the radio resource management (RRM) under advanced performance metrics such as AoI and energy-efficiency (EE) is a  challenging task.
At the same time, it has a crucial impact on the network performance, connection capacity, and cost.
The RRM via the ML algorithms considering unified network performance metric in the area of wireless communication has not been well addressed, and it is a critical issue and pivotal for future complex and heterogeneous networks as 6G. 
\\\indent
In this paper, we  devise a ML-based radio resource scheduler capturing freshness of information via formulating a stochastic optimization problem. 
The salient features of the proposed scheduler are 1) unified network performance metric 
is optimized 
 by formulating a novel constrained Markov decision problem (CMDP); and 2) solving this problem via deep neural-based  RL (DRL).
	\subsection{{ Contributions and Research Outcomes}}
	The  contributions of this paper are summarized as follows:
	\begin{itemize}
	\item	In contrast to average AoI (AAoI) minimization \cite{Infocom_OMANOMA,AI_AoI,RL_Bennis}, power minimization \cite{Moltafet_Power,EE_AOI}, and EE maximization \cite{Energy_AoI_UAV}, we study a unified network performance metric which is defined by energy spectral efficiency (ESE) and AAoI. To this end, we exploit multi carrier non-orthogonal multiple access (NOMA) in a multi-access edge computing (MEC) assisted base station (BS) and the main aim is dynamically allocate power and subcarriers to the uplink users. 
		\item 
A novel CMDP is formulated whose objective is 
		 maximizing  the long-term joint ESE and AAoI via packet transmission decision and RRM
		 taking edge-level computing and radio  resources limitations into accounts. 
	 Also, we provide a unified computing and RRM framework capturing the freshness of information. 
	\item
		We propose that each user can have multiple types/classes of information  with different levels of AAoI constraints. This makes our framework  applicable for a wide range of  applications\footnote{This is because of that	in practice, each user has multiple information with different degrees
		of importance such as security, control channel, and location  information.}.
\item 
		We develop DRL 
to solve the formulated problem to provide an intelligent scheduling algorithm. Also, we propose non-uniform channel quantization and power discretization levels to guarantee the fast convergence of the DRL and its performance.
	%
\item The effectiveness of our model and the main parameters impact (e.g., available spectrum  and packet size) on AoI and the system  reward is validated through extensive simulations. Numerical results  illustrate our work outperforms existing state of the art baselines up to  64\% in objective function and 51\% in  AAoI, which are stated as examples. Also, we observe that the proposed reward function converges well with a negligible loss. 
		\end{itemize}
	\subsection{{Paper Organization and Notations}} 
	The outline of this paper is organized as follows. System description and problem formulation are stated in Section \ref{SD_PF}.  Section \ref{Solution} presents the proposed solution. Finally, simulation level evaluation and conclusion remarks are expressed in Sections \ref{Simulation_Re} and \ref{Conclusion}, respectively. 
	\\
	\textbf{Symbol Notations:}
	$ \Bbb{N}_{+} $ indicates the set of non-negative integer numbers and $ \|\cdot\|_0 $ is $ l_0 $ norm. $\vee$ and $ \oplus $ indicate  the logical OR and XOR operators, respectively. Moreover, matrices and vectors are shown with bold uppercase and lowercase letters, respectively. Also, we indicate the value of symbol $ A $ at the beginning of time slot $ t $ as $A(t)$.
	\subsection{{Related Works}}
	We discuss the related works in two main categories: 1) AoI-based resource management; 2) ML adopted scheduling algorithms in the AoI-based optimization. 
	\subsubsection{AoI-based Resource Management} 
	Recently,  AoI in the context of resource management is studied from different perspectives \cite{Infocom, Moltafet_Power, Radio_AoI, AoI_Tony}. 
		In \cite{Radio_AoI}, an AoI-aware RRM scheme (subcarrier allocation) is devised for vehicle-to-vehicle  communications considering a network topology with the Manhattan Grid model. 
		The authors propose a proactive algorithm inspired by long short-term memory and DRL.
			Analyzing the queuing theory and stochastic geometry for the optimization of AoI is provided in \cite{AoI_Tony}. Numerical evaluations in \cite{AoI_Tony} reveal that the packets arrival rate has direct impact on the peak AoI.		

	Moreover, in \cite{Infocom}, the authors investigate the throughput of wireless networks by minimizing the weighted sum AAoI and analyze the minimum required throughput for a given AoI.   They propose different policies such as  random and   MaxWeight.  The long-term  transmit power minimization by optimizing both transmit power and subcarrier assignment under AoI and orthogonal multiple access (OMA) constraints is studied in \cite{Moltafet_Power}. To solve the formulated problem, they adopt the Lyapunov drift-plus-penalty method. 
	 Furthermore,  
	  linear weighted  combination of AoI and total energy consumption as the objective function is studied in \cite{EE_AOI}. They derive an optimal scheduling policy that finds the globally optimum point for AoI and consumed energy trade-off.  The most recent work in this area at the time of this paper is \cite{Energy_AoI_UAV}.  The main purpose of \cite{Energy_AoI_UAV} is to maximize the EE of the unmanned aerial vehicles network considering   AoI as a constraint in which maximum AAoI for each user should not violate a predefined threshold. 
	\subsubsection{ML-based Optimization in AoI Context}
Up to now, a plethora of researches has focused on  the application of ML-based solutions in the communication era, especially RL as a subset of ML. This method is very useful  in the network-level automation, tackling the complexity of the non-convex optimization problems, giving near-optimal solutions  \cite{ML_Survey, AI_AoI, AoI_RL}. 
	The authors in \cite{ML_Survey} provide a survey on the application of  ML and deep learning  as an efficient method in addressing the challenges of resource management for wide range of internet of thing  (IoT) networks. They point out some challenges in this area such as heterogeneity of data and an avalanche effect of ML.
The authors of \cite{AI_AoI} investigate the minimization of the  AAoI by formulating a partially observable Markov game in wireless Adhoc networks.
%
%
%
They adopt deep RL to find the online policy and dynamic scheduling. In \cite{AoI_RL}, the real time status monitoring system enabling energy harvesting via broadcasting signal is studied, where the authors optimize the weighted  AAoI.  Moreover, 
\cite{RL_Bennis}  studies the problem of AAoI minimization by considering the probability of
threshold violation of AoI in ultra-reliable and low latency communication. They exploit the RL approach to find
the scheduling policy from real observations.  Furthermore, the authors of \cite{Energy_AoI_UAV} adopt DRL to optimize the trajectory points for flying unmanned aerial vehicles (UAVs). They consider  EE maximization with AAoI as a constraint. 

In order to explain  the difference of our work from the existing studies clearer, we emphasize  that this is the first work which  devises a novel CMDP {which contains} a ML-based RRM and packet transmission decision. 
Here, we point out that the most similar works to our work are
	 \cite{Moltafet_Power, Radio_AoI, Energy_AoI_UAV}. 
	 	In \cite{Moltafet_Power}, the authors consider the power minimization problem in a wireless sensor network  with AAoI constraints. The authors adopt the Lyapunov optimization theory to minimize the power consumption of users while keeping the data fresh at the destination. However, they do not distinguish between the packet generation and transmission time and AoI is only dependent on the sampling action in which each user can take the sample if the considered radio resources are enough. While the data sampling action  is an independent random process.
	 	Also, compared with \cite{Radio_AoI}, besides subcarrier allocation, the power allocation and packet dropping
	 	with unified performance metric are studied in our work. Moreover, as a new work in this field, we  consider the buffer and processing capacity limitations in a expression of AoI.	As compared with both \cite{Radio_AoI, Moltafet_Power}, we adopt NOMA which plays the key role in the network ESE and AoI.
	 	
	 	In comparison with \cite{Energy_AoI_UAV}, although the authors study EE maximization under AAoI   constraint via DRL, we propose a novel maximization of joint long-term ESE and  AoI  considering AAoI as a constraint via optimizing radio access and computing resources. However, they optimize the trajectory of UAVs and do not consider radio resource limits and roles on the network performance.  
 \vspace{-1cm}
	\section{{system model description and Optimization Problems }}\label{SD_PF}
\begin{figure*}[!ht] 
		\centering
		\includegraphics[width=.65\textwidth]{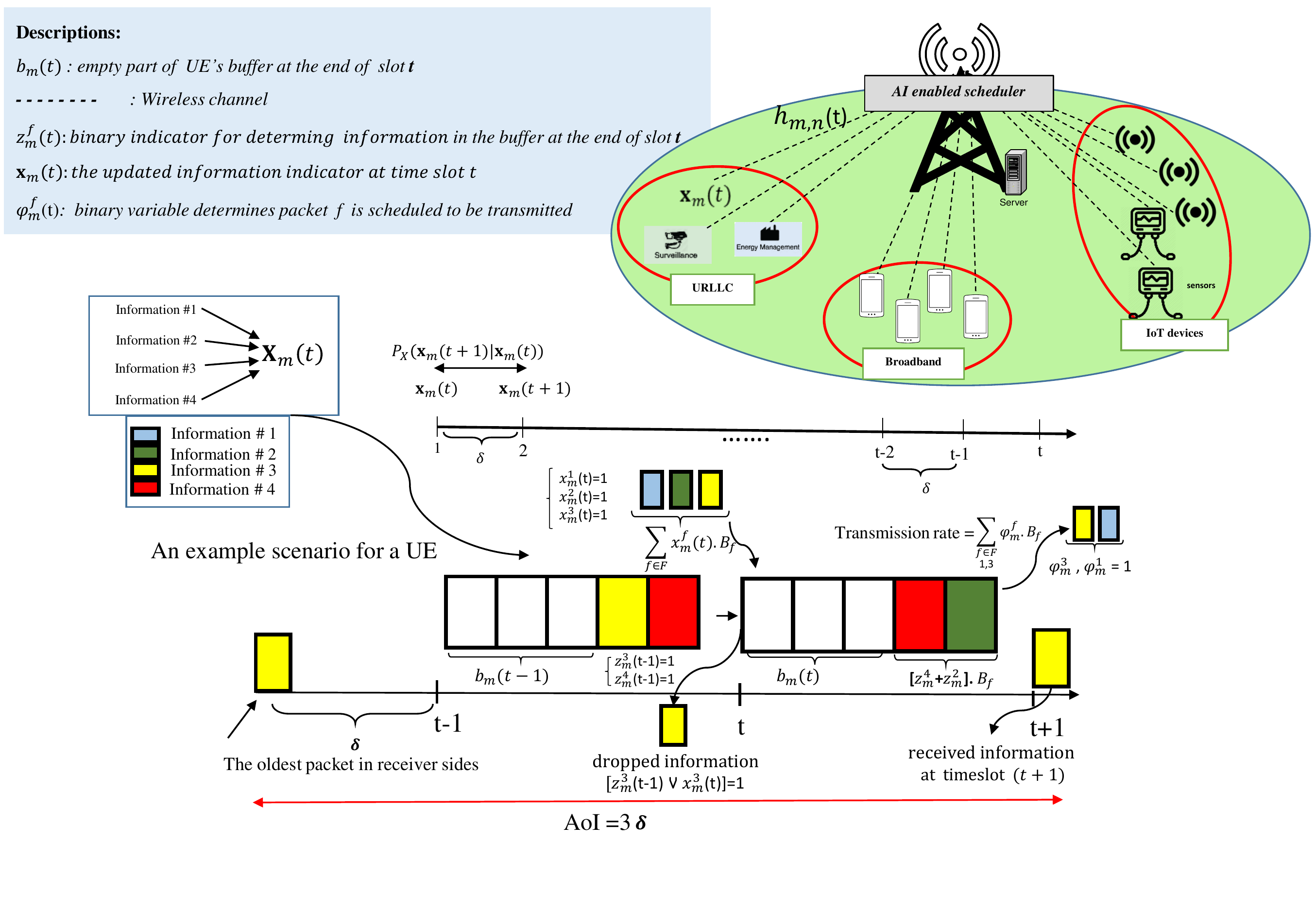}
		\vspace{-2em}
		\caption{Considered system model assuming $ 4 $ types of different information. The arbitrary scheduling time slots in sequence are denoted as $  t-2, t-1 $, and $ t $, and UE type as $ m $. As it is illustrated, at the end of time slot $ t-1 $, information types of $ 3 $ and $ 4 $ are stored in the buffer, but the oldest packet among $ 3 $ and $ 4 $ is information type $ 3 $ which is generated and received at time slot $ t-2 $. 
		} 
		\label{System_Model}
	\end{figure*}

	We consider a wireless BS enabled with a MEC server with resources indicated by
$	\bold{r}=\big[w,\alpha^{\max}\big],$
	where  $w$ and $ \alpha^{\max},$  are the total available bandwidth for uplink transmission 
	in the radio part and the maximum CPU cycles per unit time (second),  respectively. To make better understanding, a description of the symbol notations and variables are summarized in Table \ref{Table_Not}.
	\begin{table}
				\caption{Table of the main notations/parameters and variables}
		\label{Table_Not}
		\centering
	\renewcommand{\arraystretch}{.9}
	\begin{tabular}{c||l}
		\hline\textbf{Notation(s)} & \textbf{Definition} \\
		\hline  \multicolumn{2}{l}{~~~~~~~~~~~~\textbf{Notations/parameters}}
		\\
		\hline$\alpha^{\max}$ & Maximum CPU capacity of the server \\
		\hline$w/\varpi $ & Total bandwidth/subcarrier spacing of subcarrier $ n $\\
		\hline $N/\mathcal{N}/n$ & Number/set/index of subcarriers \\
		\hline$M / \mathcal{M}/m$ & Number/set/index of UEs \\
		\hline$F / \mathcal{F}/f/B_f$ & Number/set/index/size of information data type \\
		\hline$ t/\delta$ & Index/duration of each time slot 
\\	\hline	$\vartheta  $&  CPU cycles needed to process one bit of data
		\\
		\hline $x_{m}^{f}(t)/z_{m}^{f}(t)$ &The updated/buffered information $ f $ indicator for UE $ m $ at time slot $ t $ \\
		\hline $\theta_{m}^{f}(t)$ & The evolution of AoI of information  $ f $ for UE $ m $ \\
		\hline $N_0$ &Noise power
		spectral density \\
			\hline $P_{m}^{\max}$ & Maximum allowable transmit power of UE $ m $ \\
		\hline $\beta(t)/b_{m}(t)$ & Empty part of the buffer of server/UE $ m $ \\
		\hline $h_{m,n}(t)/\gamma_{m,n}(t)$ & Channel/SNR state of UE $ m $ on subcarrier $ n $ \\
		\hline \multicolumn{2}{l}{~~~~~~~~~~~~\textbf{ Optimization Variables}} 
		\\
		\hline$\varphi_{m}^{f}(t)$ & Binary transmission indicator   for information  $ f $ of UE $ m $ \\
		\hline$\rho_{m,n}(t)$ & Binary subcarrier assignment of UE $ m $ on subcarrier $ n $
		\\ \hline$p_{m,n}(t)$ &Transmit power   of UE $ m $ on subcarrier $ n $
		\\
		\hline
	\end{tabular}
	\end{table}
	The total available bandwidth, $ w $, is divided into a set $ \mathcal{N} $ of $ N $ subcarriers indexed by $ n $ with subcarrier spacing $ \varpi $.
	The considered system model is depicted in Fig. \ref{System_Model} in which different user equipment (UE) types 
 served by the BS.  Details are in the caption of Fig. \ref{System_Model}. 
	We discretize the continuous time  into  time slots, each of which is indexed by $t\in\Bbb{N}_{+}$ with the equal time duration as $\delta$. Our proposed system model is assumed to be delay free \cite{Moltafet_Power} and packet transmission scheduling is done at the beginning of each time slot.
	\vspace{-0.5cm} 
	\subsection{{UEs'  Information Updates}} 
	We consider a  set $\mathcal{M}=\{1,\dots,M\}$ where $M$ is the total number of UEs. Each UE has a limited buffer capacity denoted by $ b_{m}^{\max} $. 
	We consider a set   $ \mathcal{F}=\{1,\dots,f,\dots,F\} $ of $ F $ independent types of information for each UE,
	 each of which is encapsulated in a packet (also indexed by $ f $). Each  information 
	packet of type $ f $  has different specifications 
	and 
	%
	contains  $ B_f $ bits\footnote{Without loss of generality, we assume that all packets have the same sizes.}.
	We define a random variable $ x_{m}^{f}(t)$ in such a way that, if $ x_{m}^{f}(t)=1 $, it shows that UE $ m $ has an updated information of type $ f $, otherwise it is zero.  
	Based on this, 
	we consider that at the beginning  of  time slot $t$, UE~$m$'s updated information packets are $ \bold{x}_{m}(t)=[x_{m}^{f}(t)] $
	{and} the number of  information packets {for each user} is obtained by $ \|\bold{x}_m(t)\|_{0} $.
	The updated status information is generated randomly according to an unknown Markov process. 
	Based on {this}, the updated packets  at {each time slot}, i.e., $ \bold{x}_{m}(t)$  are updated to $ \bold{x}_{m}(t+1)$ with a transition probability as   
$	\Pr\left(\bold{x}_{m}(t)~\big|~\bold{x}_{m}(t-1)\right),~\forall m\in\mathcal{M}, t\ge 1.$
	We can set $\Pr\left(\bold{x}_{m}(t)~\big|~\bold{x}_{m}(t-1)\right)=\frac{1}{2^F} $ which follows the uniform distribution over power set of the total types of information. This approach is similar to the task generation transition probability of [Page 5 of Section IV, \cite{Privacy}].
	\subsection{Optimization Variables and Related Parameters}
	We define $\varphi_{m}^{f}(t) $ as an optimization binary variable indicating that information packet of type $ f $ user $ m $ is transmitted to the server at the beginning of time slot $t$.
We consider that  all the information packets generated at the beginning of time slot $ t $ are not necessarily transmitted to the server at the same time and therefore unsent packets will be stored in the buffer. 
Accordingly, 
	let  $ z_{m}^{f}(t) $ be a binary indicator representing the presence of information packet of type $ f $ for UE $ m $ in {its} buffer after the scheduling process at time slot $ t $ with $ z_{m}^{f}(t)=1$, otherwise $ z_{m}^{f}(t)=0$.
	Also, we define $ b_m(t) $ as the empty section of users' buffer at time slot $ t $. Furthermore, we have power allocation and subcarrier assignment variables which are discussed in Section \ref{Radio_Con}.
	\subsection{System Model's Related Constraints}
We discuss the following two system constraints: 1) buffering and computing constraints; 2) radio resource scheduling constraints. 
\subsubsection{buffering and computing constraints}
		Before introducing the  constraints, we define the dropping policy as follows: 
	\begin{Defi}
		\label{Dropping_Policy}
When an information packet of type $ f $ already exists in the buffer from the previous time slots, and a packet of the same type (information $ f $) is generated, the old packets (information in the buffer) will be replaced by the new ones.
 In this  regard, we define logical OR operation between the buffered information indicator, i.e., $ z_{m}^f(t-1) $ and the indicator of new generation, i.e., $ x_{m}^{f}(t) $. 
 Based on this, we update the buffered information by the following rule.
	\end{Defi}
	By considering the transmitted information and randomly generated/updated information packets, we can update the buffered information recursively as follows:
	\begin{align}
	&z_{m}^{f}(t)=\underbrace{\underbrace{\big[z_{m}^{f}(t-1)\vee x_{m}^{f}(t)\big]}_{\textit{Packet existence in the buffer}} \oplus \underbrace{\varphi_{m}^{f}(t)}_{\textit{Transmission part}}}_{\textit{Remaining data in the buffer}},~z_{m}^{f}(0)=0.
	\label{Z_Buffer}
	\end{align}
	$\bullet~$\textit{Buffer Capacity Constraint:}
	Based on the information packets left in the buffer, we can update the empty section of UE's buffer at each time slot by
	\begin{align}
	&b_{m}(t)=b_m^{\max}-\sum_{f\in\mathcal{F}}z_{m}^{f}(t)\cdot B_f. 
	\end{align} 
		The UEs' updated information packets, $ \bold{x}_{m}(t) $, should always be less than its buffer's empty section. Therefore, the empty part of the UE's  buffer should always be positive. 
	To this regard, we introduce the following constraint:
	\begin{align}
	&\label{UE's buffer capacity}
	b_{m}(t)\ge 0,~\forall m.
	\end{align}
	$\bullet~$\textit{{Information Transmission
			Constraint:}} {When a schedule is made to send an information packet of type $ f $ at time slot $ t $, either this packet
		must be in the buffer from previous time slots or must be generated at that time slot.}
	This is ensured by the following constraint
	\begin{align}
	\label{Tran_Con}
	&
	\varphi_{m}^{f}(t)\le \underbrace{z_{m}^{f}(t-1)}_{\textit{Already in buffer}}\vee \underbrace{x_{m}^{f}(t)}_{\textit{Generated}}.
	\end{align}
	$\bullet$~\textit{Processing Constraint:}
	The scheduled bits to run at the server should not violate the CPU cycle capacity of the server. At the same time, we assume that the information processing  on the CPU should be finished within the duration of one slot with duration $ \delta $.  Therefore, we introduce the following constraint:
	\begin{align}
	\label{CPU_Cons}
	F_i(t) \triangleq \frac{\sum_{m\in\mathcal{M}}\sum_{f\in\mathcal{F}}\varphi_{m}^{f}(t)\cdot B_f\cdot\vartheta}{\delta}\le \alpha^{\max},
	\end{align}	
	where $\vartheta  $ is the number of CPU cycles needed to process one bit which depends on the type of application, and $ F_i(t) $ is the total CPU-cycle frequency  of the server that is utilized at time slot $ t $. 
	Therefore, the
	CPU energy consumption at the time interval $ \delta $ at the server is given by \cite{Power_Delay_Globcom}
	\begin{equation}
	E^{(\mathrm{CPU})}_{\text{Server}}(t)=\underbrace{\zeta\cdot F_i^3}_{\text{Power consumption}}\cdot\delta, 
	\end{equation} 
	which $\zeta$ is the effective switched capacitance related to the
	chip architecture.
	\subsubsection{Radio Resource Scheduling Constraints}\label{Radio_Con}
	Our aim is to optimize power allocation and subcarrier assignment variables as a radio resource scheduling. First let  
	 $ h_{m,n}(t) $ be the channel coefficient between UE $ m $ and the BS on subcarrier $n$ at time slot $ t $ assumed to be perfectly known \cite{Moltafet_Power}. To multiplex users on given orthogonal subcarriers, we adopt uplink NOMA, especially power-domain NOMA. To this end, we present wireless channel model and a brief description of the working principles of uplink NOMA and relevant optimization constraints in the following.
	\\$\bullet~$\textit{Wireless Channel Model and States:}
In this paper, our channel model, i.e., $ h_{m,n}(t) $ is as follows. We produce channel coefficient and channel states in two parts. We model  $ h_{m,n}(t)=g_{m,n}(t)\cdot\sqrt{(\frac{d_0}{d_{m}})^{3}} $, where $ g_{m,n}(t) $ indicates the small scale fading and is generated according to complex  Gaussian distribution with zero mean and variance $ \sigma^2 $, i.e., $ \mathcal{CN}(0,\sigma^2) $ and $ d_0 $ is the reference distance. Also, $ d_m $ is the geometric distance between the BS and UE $ m $.  
\\\indent We generate  $ g_{m,n}(t) $  over time slots based on independent identical distribution (i.i.d),
after that, we transform its values into   channel states. To this end, we quantize  the value of channel gain, i.e.,  $ |h_{m,n}(t)|^{2} $  into finite regions as a channel states level assuming the number of regions is $ L $. 
Based on this, $ |h_{m,n}(t)|^{2} $ in each time slot $ t $ can be mapped into a region from all states set $\mathcal{H}$, where $\mathcal{H}=\left\{l_1,\dots,l_{L}\right\}$. In addition, each state $ l_i, i\in\{1,\dots,L\} $  corresponds to a region denoted by  $ [a_i,a_{i+1}) $ with $ a_1=0<a_2<\dots<a_{L+1}=\infty $. It is worthwhile to mention that the precision of channel quantization highly depends on $ L $. According to Rayleigh distribution function\footnote{Large values are in some intervals (around mean of distribution) and lower values in the other intervals}, we propose a non-uniform quantization to enhance the accuracy.    
By this  model, we reduce the state space from infinite possible states to the limited number of states levels to improve   convergence speed and accuracy.
\\$\bullet~$\textit{Basic Principles of Uplink NOMA:} In general, three principles exist in an uplink NOMA: 1) \textit{subcarrier assignment}, 2) \textit{power allocation}, and 3) \textit{applying successive interference cancellation (SIC)}. 
\\\indent
In \textit{subcarrier assignment principle}, based on NOMA, each subcarrier $n$ can be assigned to no more than $L_n$ UEs in the converge of the BS, simultaneously. Therefore,  we introduce the following  subcarrier assignment constraint:
	\begin{align}
	\label{Spectrum_Resource_Con}
	&
	\sum_{m\in\mathcal{M}}\rho_{m,n}(t)\le L_n,\forall n,
	\end{align}
	where  $  \rho_{m,n}(t)\in\{0,1\} $ is the subcarrier assignment variable of  subcarrier $ n $ for UE $ m $.
	In \textit{power allocation principle}, each UE $ m $ transmits its signal on subcarrier $ n $ with power allocation variable $ p_{m,n}(t) $ at each time slot $ t $ in which its value is limited by UE's maximum allowable transmit  power denoted by $ P^{\max}_{m} $. Clearly, this imposes the following power allocation constraint:
	\begin{align}
		\label{Power_Con}
	&
	\sum_{n\in\mathcal{N}}\rho_{m,n}(t)\cdot p_{m,n}(t)\le P^{\max}_{m}, \forall m. 
\end{align}
Since in NOMA, distinct signals of UEs are superposed on a subcarrier, the intra-cell interference becomes vulnerable on NOMA performance (both downlink and uplink). In this regard, SIC has proposed to guarantee NOMA performance compared to OMA \cite{SIC_Ekram}. Based on \textit{SIC principle} in uplink NOMA, the BS successively decodes and cancels the signals of strong channel gain UEs, before decoding the signals of weak channel gain UEs \cite{SIC_Ekram}.  It is worth noting that as discussed in [Section B on Page 5 of \cite{SIC_Ekram}], SIC ordering based on the received signal power can be approximated by ordering of UEs based on channel gain. 
 Note that in the uplink transmission, the signal of each UE is transmitted over distinct channel gain and the BS receives signals of all UEs over their channels, simultaneously. Therefore, UEs with strong channel gain has the highest order in SIC ordering at the BS and UEs with the weak channel gain produce interference to the UEs which have strong channel gain. 
Based on these, the received  {signal-to-noise at the BS side} of UE $ m $ on subcarrier $ n $	which is denoted by  $ \gamma_{m,n}(t) $  is calculated by \eqref{SINR} shown at the top of next page,
	\begin{figure*}
	\begin{align}
\label{SINR}
	\gamma_{m,n}(t)=
	\frac{p_{m,n}(t)\cdot\left|h_{m,n}(t)\right|^{2}}{\varpi \cdot N_{0}+\sum \limits_{m'\in\left\{\mathcal{M}\backslash m,~\Big|~\left|h_{m',n}(t)\right|^{2}\le\left|h_{m,n}(t)\right|^{2}\right\}}\rho_{m',n}(t)\cdot p_{m',n}(t)\cdot \left|h_{m',n}(t)\right|^{2}},
	\end{align} 
			\hrule
\end{figure*}
	where $ N_0 $ is the additive noise power spectral density. Therefore, according to the  Shannon’s capacity formula, the data rate (bits per second) of UE $ m $ on all assigned subcarriers at  each scheduling time slot $ t $ is obtained by 
\begin{align}
r_{m}(t)=\sum_{n\in\mathcal{N}}\varpi\cdot \rho_{m,n}(t)\cdot \log _{2}\left(1+\gamma_{m,n}(t)\right).
\end{align}
Moreover, the  packets that are determined (via $ \varphi_{m}(t) $) to be sent to the BS/server side should be encapsulated on the radio barriers. To satisfy that the uplink data rate of each UE $ m $ supports the transmit data rate in time duration $ \delta $, we introduce the following constraint:
\begin{align}\label{Rate_Con}
\sum_{f\in\mathcal{F}}\varphi_{m}^{f}(t)\cdot B_{f}\le r_m(t)\cdot \delta, \forall m.
\end{align}
	\subsection{Modeling of AoI}
	%
	Our system model is delay free (does not include propagation, transmission, and queuing delays) and packet transmission scheduling is done at the beginning of each time slot, therefore, when $ \varphi_{m}^{f}(t)=1 $, information $ f $ will be received at the destination within that time slot with no delay.
		We define $ \theta_{m}^{f}(t) $ as the AoI of packet of type $ f $ for UE $ m $ at the beginning of time slot $ t $ as
	the time elapsed since the generation of the last successfully received. 
	 Since at $ t=0 $, the server has no updated received packet information, we set $\theta_{m}^{f}(0)=\infty$.  It is worthwhile to mention that in our scenario, AoI is the number of time slots elapsed  since the generation of the most up-to-date information packet of type  $ f $ successfully  received at the server.
	Upon reception of an information packet, the AoI will reset to the delay  that the packet experiences going through the transmission system. 
		Therefore, we can categorize the evolution of AoI into three possible states for UE $ m $ and information $ f $ as follows:
		\\
		 1) The information is  generated and transmitted at time slot $ t $, simultaneously, which means mathematically we have  $ x_m^f(t)=1 $ and   $ \varphi_m^f(t)=1 $. In this case, AoI is $ \delta $. 
		 \\ 2) The information is generated at time slot $ t $, but is not transmitted at  this time and stored in the UE's buffer which means mathematically we have   $ x_m^f(t)=1 $ and   $ \varphi_m^f(t)=0 $. In this case, AoI is increased by step $ \delta $.
		 \\3) The information is not generated at time slot $ t $, but transmitted in this time which means mathematically we have   $ x_m^f(t)=0 $ and   $ \varphi_m^f(t)=1 $. This implies that the information is generated in the previous time slots as $ t'\in \mathcal{T}' $ where $ \mathcal{T}' $ can be a set as $\mathcal{T}'=\{1,2,\dots,t-2,t-1\}$. Moreover, in this state (\# 3), it is possible that the information is generated multiple times in $ \mathcal{T}' $. In such way, based on defined dropping policy ({definition \ref{Dropping_Policy}} and Eq. \eqref{Z_Buffer}), the old information is dropped and freshest information is stored in the buffer which is indicated by $ z_{m}^f(t') $. In this state, the AoI  resets 
		 to the 
		 time duration between its generation and most recently reception. In order to find the last generated time of the packet of type $ f $,
		 we can utilize backward propagation by values of $ z_m^f(t) $ and $ x_m^f(t) $ as follows. Note that since in this case, the AoI has significant difference from the other two states, we denote the AoI by $  \mu_{m}^{f}(t) $ obtained by \eqref{backward}
		 assuming that its  initial value is $  \mu_{m}^{f}(0)=\infty$.
		\begin{align}
	\mu_{m}^{f}(t)=\begin{cases}
	\mu_{m}^{f}(t-1)+\delta,& x_m^f(t-1)=0~\&~z_m^f(t-1)=1, \\
	\delta,&x_m^f(t)=1~\&~z_m^f(t)=1,
	\end{cases},
	\label{backward}
	\end{align}
	Based on these discussions on three possible states for the AoI, we formulate the evolution of AoI for each UE $ m $ and information type $ f $ as follows:
		\begin{align}\theta_{m}^{f}(t)= 
	\begin{cases}{\delta} & {, \text { if } \varphi_{m}^{f}(t)=1}~ \& ~x_m^f(t)=1, 
	\\{\mu_{m}^{f}(t)} & {, \text { if } \varphi_{m}^{f}(t)=1} ~\&~ x_m^f(t)=0, 
	\\ {\theta_{m}^{f}(t-1)+\delta} & {, \text { if } \varphi_{m}^{f}(t)=0}. \end{cases} 
	\label{AoIcal}
	\end{align}
	\section{{Scheduling Policy and Problem Formulation}}
	We aim to find the resource scheduling in terms of subcarrier, power, and packet transmission  policy based on a certain objective function and optimization constraints    which are managed and orchestrated by the radio domain slice orchestrator. 
	
\subsection{	{Proposed Objective Function as Reward Function}}
	To study the proposed network performance metrics  (e.g., ESE and AoI) deeply, we propose a novel unified metric defined by
	 the  network ESE over AAoI which is mathematically formulated as follows:
	 \begin{align}
	 \label{Main_Op}
	 	\Phi(t)=\frac{ESE(t)}{AAoI(t)} = \frac{R_{\text{Total}}(t)~}{AAoI(t)\times P_{\text{Total}}(t)},~~
	 \end{align}
	 where $R_{\text{Total}}(t)=\sum\limits_{m\in\mathcal{M}}r_m(t)~$, $ P_{\text{Total}}(t)=\sum\limits_{m\in\mathcal{M}}[\sum_{n\in\mathcal{N}}p_{m,n}(t)+P_{\text{ Circuit}}]~~$, and $ \text{AAoI}(t)=\frac{1}{M\times F}\sum\limits_{m\in\mathcal{M}}\sum\limits_{f\in\mathcal{F}}\theta_m^f(t) $ are the total instantaneous throughput,  power consumption, and AAoI, respectively. As observed from the form of \eqref{Main_Op}, our objective minimizes AAoI and maximizes ESE, simultaneously.
	 It is worthwhile to note that the meaning of function \eqref{Main_Op} can be defined as the maximum achievable value of ESE per unit of AoI.
	\\\indent
	Our main aim is to maximize the averaged long-term ESE over AoI denoted by $\bar{	\Phi}(\cdot) $ and obtained by 
	\begin{equation}\label{Long_Term_Obj}
	\bar{	\Phi}\Big({\gamma(t), \pi, \mathcal{S}_{0}}\Big) \triangleq \lim_{T\rightarrow\infty}\frac{1}{T}\cdot \mathbb{E}_{\pi}\left[\sum_{t=1}^{T}  \gamma^{t}\cdot \Phi(t) \Bigg| \mathcal{S}({0})\right],
	\end{equation}
	where $\mathbb{E}_{\pi}$ means taking expectation according to the law induced by the policy 
	$\pi(\mathcal{S}(t))$ for each state at time slot $ t $ indicated by $ \mathcal{S}(t)$;$~ \mathcal{S}({0})$ is the initial system state and $\gamma \in[0,1)$ is the discount factor; intuitively, a larger $\gamma$ means that the UE is planning for a longer future reward \cite{Privacy}. Note that the inputs of the policy (i.e., designed scheduler algorithm) are network state in each time slot $ t $ and outputs of the policy are $ \Big(\boldsymbol{\rho}(t),\bold{p}(t),\boldsymbol{\varphi}(t)\Big) $, where $\boldsymbol{\rho}(t)=[\rho_{m,n}(t)]$, $\bold{p}(t)=[p_{m,n}(t)]$, and  $\boldsymbol{\varphi}(t)=[\varphi_m^{f}(t)]$.
	\begin{Defi}
All the parameters related to a complete description of our network environment including UEs, wireless channel, and the server are defined as the system/network states.
Mathematical details are presented next. 
	\end{Defi}	
\begin{Defi}
	\label{Policy_Dif}
We define the state-action mapping method related to the optimization variables as a policy, i.e., $\pi\big(\mathcal{S}(t) \big)$. Based on this, we can define three types of polices as 1) \underbar{possible policy:}  any action from all-action space can be taken,  2) \underbar{feasible policy:}  actions who ensure constrains of \eqref{optimization_problem_M},  and 3) \underbar{optimal policy:}	 actions who ensure constrains of \eqref{optimization_problem_M} 
and maximize (global maximization) the objective function \eqref{Long_Term_Obj} can be taken. 
\end{Defi}
	\subsection{Problem Formulation}
	Based on these definitions, our aim is to find the best resource orchestration policy by solving the following optimization problem:
	\begin{equation}\label{optimization_problem_M} 
	\renewcommand{\arraystretch}{1.5}
	\begin{array}{ll}
	\mathop{\textbf{{Maximize}}}&\bar{\Phi}\Big({\gamma(t), \pi, \mathcal{S}_{0}}\Big)
	\\~~~~~
	\textbf{s.t.} 
	&  
	\lim_{T\rightarrow\infty}\frac{1}{T}\cdot\Bbb{E}_{\pi}\left[\theta_{m}^{f}(t)\right]\le \kappa_{m,f}^{\min},~\forall m, f,
	\\& 
	\eqref{Tran_Con}, \eqref{UE's buffer capacity}, \eqref{CPU_Cons}, \eqref{Rate_Con}, \eqref{Spectrum_Resource_Con}, \eqref{Power_Con}, 
	\end{array}
	\end{equation}
	where $ \kappa_{m,f}^{\min} $ is the minimum average of AoI of UE $ m $ and information type $ f $. 
	\section{{Proposed Solution}}\label{Solution}
	The proposed optimization problem \eqref{optimization_problem_M} is a long-term optimization CMDP. We utilize  \textit{Q}-learning and DRL methods which are subsets of  ML  to solve our problem. 
	Based on this, we first find the policy at each scheduling time slot $ t $  with respect to the state on each slot.
	 	Then, we take the expectation of the per-slot instantaneous reward function which is obtained by the  scheduled policy regarding 
		 \eqref{optimization_problem_M}  \cite{Privacy, Privacy_TWC}. Notably, the considered reward function is the objective function defined by \eqref{Long_Term_Obj}.
		 \subsubsection{\textit{Q}-Learning} 
	 Let  
	$ \mathcal{S}_{m}(t)=\left[h_{m,n}(t),\bold{x}_m(t),b_{m}(t),\bold{z}_{m}(t),\boldsymbol{\theta}_{m}(t)\right],$
  denote the local state of each UE $ m $ at scheduling time $ t $, where $ \bold{z}_{m}(t)=[{z}_{m}^{f}(t)] $ and $ \boldsymbol{\theta}_{m}(t)=[{\theta}_{m}^{f}(t)] $. Moreover, $\mathcal{S}_{m}(t)\in\mathcal{S}_{\text{Total}}\triangleq \mathcal{S}_{1}(t)\times\dots\times\mathcal{S}_{m}(t)\dots\times\mathcal{S}_{M}(t)= \mathcal{H}\times\mathcal{X}\times \mathcal{B}\times \mathcal{Z}\times\Theta$, where $ \mathcal{H}$ is the wireless channel states,  $\mathcal{X}\subset 2^{F}$, $ \mathcal{B}$ is the buffer sate, $\mathcal{Z}\subset 2^{F}$ is the buffered information state, and $\Theta $ is the AoI state.
	    Based on the current state $ \mathcal{S}(t)\in\mathcal{S}_{\text{Total}} $, the scheduling policy $ \pi(t) $ determines an action $\mathcal{A}(t)=\pi\Big(\mathcal{S}(t) \Big)$ 
	  	such that  $ \mathcal{A}(t)\in\mathcal{A}_{\text{Total}} $, where $ \mathcal{A}_{\text{Total}} $ is all of the feasible actions that can be taken and is defined as follows:
	  		\begin{align}
	  		\label{Action_Spapce}
	  		\mathcal{A}_{\text{Total}}(t)
	  		\triangleq\mathcal{A}\left(\boldsymbol{\rho}(t)\right)\times\mathcal{A}\left(\boldsymbol{p}(t)\right)\times \mathcal{A}(\boldsymbol{\varphi}(t)).
	  		\end{align}
	  
It can be readily noticed that, in \eqref{Action_Spapce}, all terms are obtained by \eqref{Action_Policy} on top of Page 16.
	 \begin{figure*}
 \begin{align}
 \small
 \renewcommand{\arraystretch}{1.9}
 \label{Action_Policy}
 	\begin{array}{ll}
  &\mathcal{A}(\boldsymbol{\rho}(t))= A^{\text{Sub.}}_{1,1}(t)\times \dots \times A^{\text{Sub.}}_{M,N}(t), \text{with}~
	 A^{\text{Sub.}}_{i,j}(t)=
	\left\{\{0,1\}~\big|~\eqref{Rate_Con}, \eqref{Spectrum_Resource_Con}, \eqref{Power_Con}~ \text{are satisfied}\right\},
	\\&
	\mathcal{A}(\boldsymbol{p}(t))=A^{\text{Power}}_{1,1}(t)\times \dots \times A^{\text{Power}}_{M,N}(t), \text{with}~	
A^{\text{Power}}_{i,j}(t)=\left\{\{p_1,\dots,p_L\}~\big|~\eqref{Rate_Con}, \eqref{Spectrum_Resource_Con}, \eqref{Power_Con}~ \text{are satisfied}\right\},
	\\&
	\mathcal{A}\left(\boldsymbol{\varphi}(t)\right)=A^{\text{PT}}_{1,1}(t)\times \dots \times A^{\text{PT}}_{M,F}(t), \text{with}~
	A^{\text{PT}}_{i,j}(t)=\left\{\{0,1\}~\big|~\eqref{Tran_Con}, \eqref{UE's buffer capacity}, 
\eqref{Rate_Con},~ \text{are satisfied}\right\},
	\end{array}
\end{align} 
\hrule
\end{figure*}
%
	Moreover, $ \{p_1,\dots,p_q\} $ is the discrete set of power levels and can be set as $ p_{i+1}=p_{i}+\Delta,~i\in\{0,\dots,q\},~p_0=0 $, where $ \Delta=\frac{P_m^{\max}}{q+1} $.
	The globally optimal scheduling policy is a policy which maximizes  the reward function over the possible actions.  
	\begin{remark}
		We have different values for the maximum transmit power and power allocation action mainly depending on the feasible set of \eqref{Power_Con}. At the same time, the accuracy (algorithm performance) and convergence speed of DRL highly depends on the power action levels, i.e., $ q $. Therefore, we consider different number of levels for different values of $ P_m^{\max} $, i.e., large value of $ q $ for large values of $ P_m^{\max} $.
	\end{remark}
\vspace{-2em}
	An optimal policy denoted by $ \pi^{\star} $ is obtained by solving Bellman's equation  given by \eqref{Bell_Man} on top of Page 16, where
		$  \Pr\left(\mathcal{S}^{\prime} \Big| \mathcal{S}{(0)}, \pi(\mathcal{S}(0))\right)$ indicates the 
		 transition probability from state $ \mathcal{S}(0) $ to state $ \mathcal{S}^{\prime}$ under policy (taken actions) $ \pi(\mathcal{S}(0)) $ which is taken over initial  state, i.e., $ \mathcal{S}(0) $.
\begin{figure*}
\begin{equation}
\label{Bell_Man}
\begin{aligned}
V^{\pi}(\mathcal{S}(0)) 
&=\max_{\pi}\Bbb{E}_{\pi}\left[\sum_{t=1}^{\infty} \gamma^t\cdot
\Phi\Big(\mathcal{S}{(t)}, \pi\left(\mathcal{S}{(t)}\right)\Big) 
\Big| \mathcal{S}{(0)}\right] 
\\
&
=\max_{\pi}\Bbb{E}_{\pi}\left[\Phi\Big(\mathcal{S}{(0)}, \pi(\mathcal{S}{(0)})\Big)
+ \sum_{\mathcal{S}^{\prime} \in \mathcal{S}_{\text{Total}}}\gamma^t \Pr\left(\mathcal{S}^{\prime} \Big| \mathcal{S}{(0)}, \pi(\mathcal{S}(0))\right) \cdot V^{\pi}\left(\mathcal{S}^{\prime}\right)\right],
\end{aligned}
\end{equation} 
\hrule
\end{figure*}
		Based on the dynamic programming \eqref{Bell_Man}, the objective function, i.e., the reward function can be admitted into state-action-value \textit{Q}-function which is given by \eqref{Q-function} (on top of Page 8) \cite{Radio_AoI}, and $ \zeta(t)\in[0,1) $ is the learning rate \cite{RL_Book}.
		 Based on this Q-value is updated according to the temporal-difference
		update.	
		\begin{figure*}
		\begin{equation}
		\small
			Q\left(s{(t)}, a{(t)}\right)
 \leftarrow 
	Q\left(s{(t)}, a{(t)}\right)+\zeta(t)\Big\{\Phi(t)+\gamma \underset{a(t+1)\in\mathcal{A}(t+1)}{\arg \max}\Big[Q\left(s{(t+1)}, a(t+1)\right)-Q\left(s{(t)}, a{(t)}\right)\Big]\Big\},
\label{Q-function}
		\end{equation}
		\hrule
	\end{figure*}
To reach convergence of the learning process, the following standard conditions should be met,
$\sum_{t=1}^{\infty} \zeta(t)=\infty,~~
\sum_{t=1}^{\infty}\left[\zeta(t)\right]^{2}<\infty$ \cite{RL_Book}.
\\\indent
	According to the  transformed dynamic programming \eqref{Q-function}, we can use two timescale learning algorithms, namely \textit{exploration} and \textit{exploitation}, i.e., $ \epsilon $-greedy algorithm \cite{RL_Book, Privacy, RL-Signal-Processing}, to obtain an optimal policy which is described as follows. 
1) \textit{Exploitation:} In each time slot $ t $, in which the  network state is described by $ \mathcal{S}(t) $, the scheduler observes the current state, and then, takes an action  such that $ a(t)=\arg\max_{a\in\mathcal{A}(t)}Q[s(t),a] $ based on the current estimate $Q[s(t),a(t)]$ with probability $ [1-\epsilon(t)] $ where $\epsilon(t) $ decreases according exponential function of $ t $. 2) \textit{Exploration:} The scheduler takes an action randomly (can be set uniformly \cite{Privacy}) with probability $ \epsilon(t) $   without considering  \textit{Q}-function value.  The main steps  of the adopted RL algorithm with model-free   \textit{Q}-learning is summarized in Al. \ref{AL_RL_Policy}.
\begin{algorithm}
	\renewcommand{\arraystretch}{0.7}
	\caption{Resource scheduling policy}
	\small
	\label{AL_RL_Policy}
	\DontPrintSemicolon
	\KwInput
	{$ t=0, s(0)=s_0, Q[s(0),a(0)]={0}$}
		\Repeat{The learning process till convergence}
		{
	{Observe the current state $ \mathcal{S}(t) $}
	
	{Take  action $ a(t) $ as a current policy as follows: ~~~~~~~~~~~~~~~~~~~~~~~~~~~~~~~~~~~~~~~~~~~~~~~~~~~~~~~~~~~~~~~~~~~~~~~~~~~~~~~~~~~~~~~~~~~~~~~~~~~~~~~~~~~~~~~~~~~~
		$\bullet$ Randomly with probability $\epsilon $ ~~~~~~~~~~~~~~~~~~~~~~~~~~~~~~~~~~~~~~~~~~~~~~~~~~~~~~~~~~~~~~~~~~~~~~~~~~~~~~~~~~~~~~~~~~~~~~~~~~~~~~~~~~~~~~~~~~~~~~~~~~~~~~~~~~~~~~~~~~~~~~~~~~~~~~~~~~~~~~~~~
		$\bullet$ Otherwise, $a(t)=\arg\max_{a\in\mathcal{A}(t)}Q[s(t),a]$
}
	
	{\textbf{Update}
		
	$\bullet$ Network states,
	$\bullet$ ESE and AoI,
	$\bullet$ \textit{Q}-function
	
}

{$ t\leftarrow t+1 $}
}

{}
		\KwOutput{$\pi$}
\end{algorithm}
\vspace{-2em}
\subsubsection{\textbf{Deep Q Networks (DQN)}}
To reach the fast convergence capability of the proposed \textit{Q}-learning algorithm for large sizes of states and actions, we adopt DRL. Moreover, DRL has gains in both experience replay and  network cloning. DRL is based on deep neural networks (DNN)  where the \textit{Q}-function is approximated by a double deep \textit{Q}-network \cite{DRL}. To do so, \textit{Q}-function is approximated as 
$Q\left(\boldsymbol\rho, \boldsymbol{\varphi}, \bold{p}\right) \approx Q\left(\boldsymbol\rho, \boldsymbol{\varphi}, \bold{p} ; \boldsymbol{\Xi}\right),$ 
	 where  $\boldsymbol{\Xi}$ is the vector of the wights of the neural (deep neural) network function approximator as stated in \cite{DRL}. In the RL community, the linear approximtor is a typical estimator \cite{DRL}. 
	  DNN can be trained by adjusting the parameters $\boldsymbol{\Xi}(t)$ at each time slot $ t $ to reduce the mean-squared error in the Bellman equation \cite{DRL}. 
	  \\\indent To improve the stability \cite{DRL} of our DQN, we use the ``reply memory strategy'' \cite{DRL,Radio_AoI} which stores the limited number of past experiences on a memory. Past experiences includes state-action pairs and achieved cost.
Therefore,  in learning/training process, instead of using only the current experience, the neural network can be trained via sampling mini-batches of experiences from replay memory  uniformly at random. This strategy reduces the correlation between the training examples and ensures that
the optimal policy is not  driven to a local minima.
	\vspace{-2em}
	\section{Performance Evaluation}\label{Simulation_Re}
This section presents the numerical results that investigate the proposed system model and measure its performance against conventional baselines. We provide numerical analysis to investigate the behavior of the proposed reward function and the impact of different parameters.  
	\subsection{Simulation Environment}
	We consider a BS with a single computing server  located at the center of a
	service coverage area and users are randomly
	distributed in the cell with a radius of $ 500 $ m.
	Moreover,  the small scale fading of the wireless channels is generated based on the 
	complex Gaussian distribution with zero mean and variance $ 1 $, and large scale is based on path loss  and all
	users perform random moving with a speed of $ v = 10 ~m/s $.
	For the DQN framework, each network of all DQN units has
	3 hidden layers with 64 neurons per layer. During the training of DNNs, we adopt the rectified linear unit (ReLU)
	function as the activation function. The simulation results are obtained in Pytorch version 1.4.0 as a Python ML library.  
	Other simulation parameters are based on the values stated in Tables \ref{Sim_Con_Par} and \ref{Sim_DQN_Par}.
	\begin{table}[t]
		\centering
		\caption{System parameters }
		\label{Sim_Con_Par}
			\renewcommand{\arraystretch}{.7}
		\begin{tabular}{c||c}
			\hline \textbf{Parameter} & \textbf{Value} \\
			\hline Number of UEs/subcarriers & $5/[2, 3]$\\
			\hline Number of information types ${F}$ & $5$ \\
			\hline Subcarrier spacing & $60~ \mathrm{KHz}$ \\
			\hline Noise power spectral density $\sigma^{2}$ & $-174 \mathrm{dBm} / \mathrm{Hz}$ \\
			\hline Scheduling slot duration $\delta$ & $0.01$ second \\
			\hline Packet size $\beta_f$ & [500, 1000, 1500, 2000] bits \\
			\hline Maximum transmit power $P_m^{\mathrm{max}}$ & [-10, 0, 10, 20, 30] dBm \\
			\hline Circuit  power per UE $P_{\mathrm{Circuit}}$ & 0.2 Watts \\
			\hline CPU cycles per bit $\vartheta$ & 737.5 \\
			\hline CPU-cycle frequency $\alpha^{\max}$ & $2 ~\mathrm{GHz}$ \\
			\hline Effective switched capacitance $\zeta$ & $2.5 \times 10^{-28}$ \\
			\hline
		\end{tabular}
	\end{table}

	\begin{table}[t]
	\centering
	\caption{Hyperparameters of DQN} 
	\label{Sim_DQN_Par}
		\renewcommand{\arraystretch}{.7}
	\begin{tabular}{c||c}
		\hline \textbf{Parameter} & \textbf{Value} \\
		\hline Number of episodes & $400$\\
		\hline Number of iterations per episode & $300$ \\
		\hline Replay memory size $ \mathcal{D} $ & $200$ \\
		\hline Mini-batch size & $32$ \\
		\hline initial $ \epsilon $ & $0.9$ \\
		\hline $ \epsilon-$greedy decrement $ \epsilon $ & $0.0001$ \\
		\hline Minimum $ \epsilon $ & $0.0001$ \\
		\hline learning rate $ \alpha $ & $0.01$ \\
		\hline Discount factor & 0.9 \\
		\hline Optimizer & Adam \\
		\hline Activation function & ReLU \\
		\hline
	\end{tabular}
\end{table}
%
%
\vspace{-2em}
\subsection{Simulation Results and Discussions}
\begin{figure}[t]
	\begin{center}$
		\begin{array}{cc}
		\subfigure[h][Convergence behavior of the reward function]{
			\includegraphics[width=0.45\textwidth]{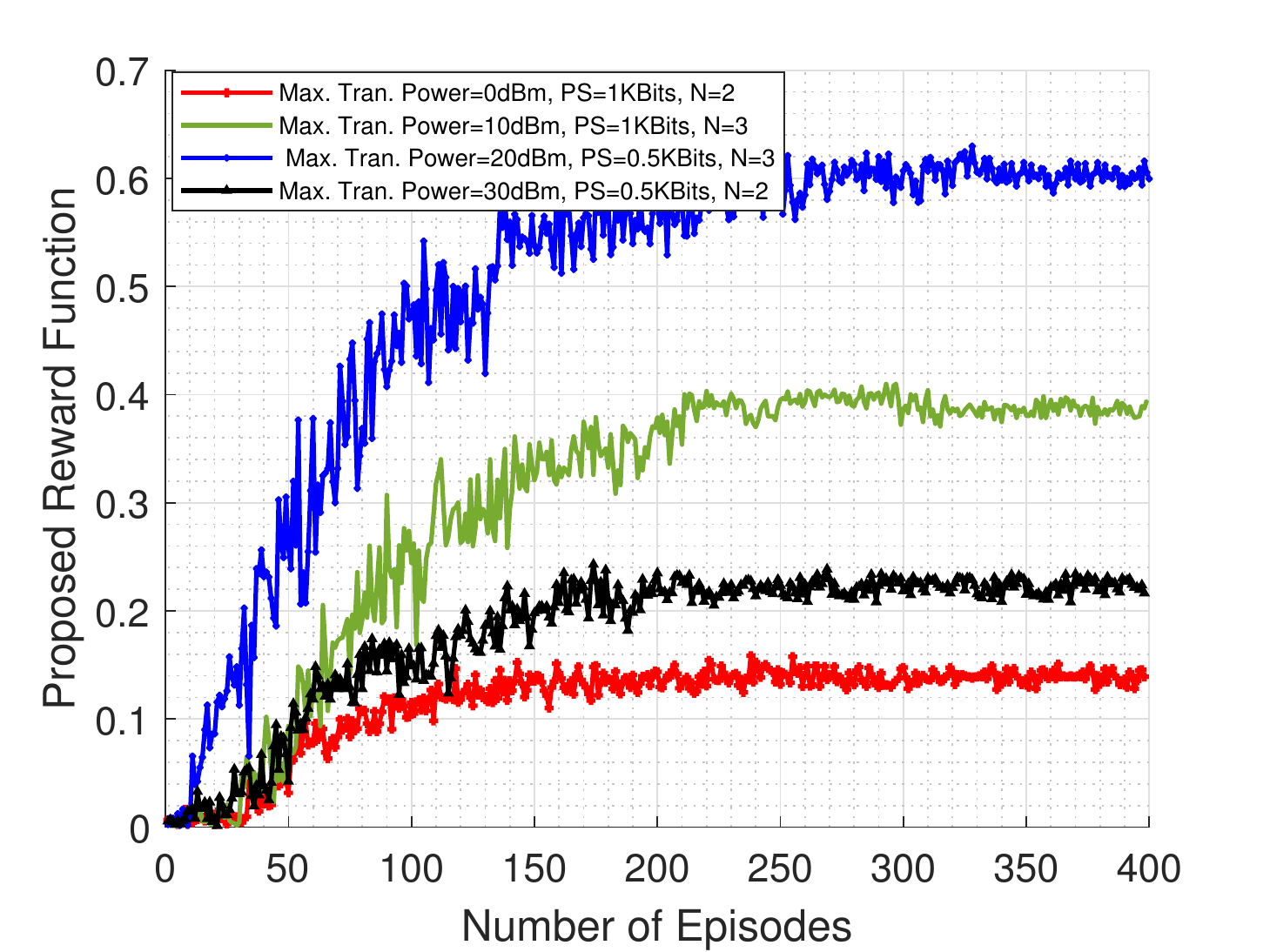}
			\label{Con_Reward}}
		\subfigure[Loss value]
		{
			\includegraphics[width=0.54\textwidth]{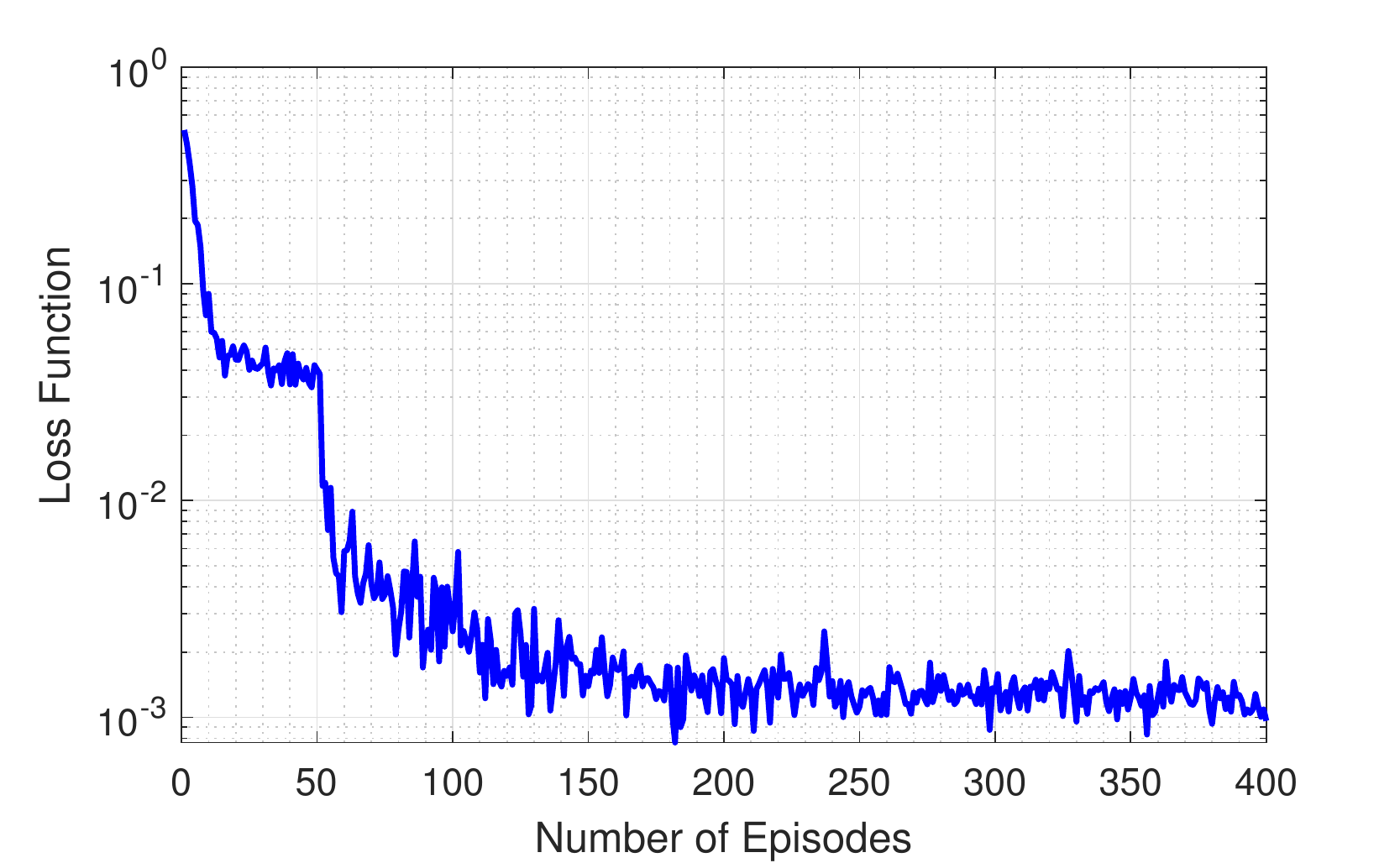}
			\label{Loss}}
		%
		\end{array}$
	\end{center}
	\vspace{-1em}
	\caption{Evaluation of convergence of the reward function and value of loss function  which is the difference  between the target network and the designed network. }
\end{figure}

This subsection discusses  the results achieved within the following schemes:
\begin{enumerate}
\item
Optimizing power allocation, subcarrier, and packet transmission scheduling in a multi carrier NOMA system via the proposed DQN as (\textbf{Proposed algorithm})
\item  Optimizing power allocation, subcarrier assignment, and packet transmission decision via proposed DQN in a OMA framework (\textbf{Baseline 1}): The OMA scheme in the context of AoI is exploited in \cite{Moltafet_Power,Infocom_1,Radio_AoI}. Besides, the performance of OMA and NOMA  are studied in such context in \cite{Infocom_OMANOMA,NOMA_AoI}. In the comparison of the state of the art \cite{Infocom_OMANOMA,NOMA_AoI}, although they provide some insights about the performance gap between NOMA and OMA on the total AAoI, we aim to provide performance comparison between them in the proposed reward function from the RRM perspective.

\item Optimizing power allocation and packet transmission decision via the adopted DQN and applying matching algorithm for the  subcarrier assignment  (\textbf{Baseline 2}):
The adopted two-sided many-to-many matching algorithm  for the subcarrier assignment is widely  exploited in wireless  communications  \cite{Zakeri1,Matching_1,Matching_3}. The main properties of the adopted matching algorithm  are as follows:
\begin{itemize}
	\item 
Users and subcarrier assumed are  selfish players (\textit{two-sided matching}). Each player generates its  own preference list on the elements of the other set
  based on matching criteria. We consider channel gain as a preference criteria in which subcarrier with higher channel gain has high priority in the preference list of users and visa versa. That means each user sorts subcarriers in the descending order with respect to the experienced channel gain.  
\item   Each subcarrier  can be assigned to at most $ L_n$ users (constraint \eqref{Spectrum_Resource_Con}) while each user can be matched with more than
one subcarrier, i.e., \textit{many-to-many matching algorithm}. 
\item The adopted matching algorithm converges to a stable matching. Also the algorithm  converges after a few iterations.  Proof of these can be found in \cite{Zakeri1, Matching_1}. 
\end{itemize}
\item Optimizing subcarrier assignment and packet transmission decision via adopted DQN and uniform power allocation  (\textbf{Baseline 3}): This baseline is adopted to show that there is no efficient power control. This basic power allocation algorithm can be exploited for power allocation \cite{Radio_AoI,PUshing}. 
\item Optimizing power allocation and subcarrier assignment  via the adopted DQN and randomly packet transmission scheduling  (\textbf{Baseline 4}). 
\end{enumerate}
\indent
Before delving into the performance comparison on the mentioned scenarios, we investigate the convergence behavior and loss value of the proposed objective function, i.e., reward function. Fig. \ref{Con_Reward} illustrates the instantaneous  reward function versus the number of episodes for different parameters. 
	From  the figure, the reward function converges and the convergence rate highly depends on the action space size. For example the  blue  curve (number of subcarriers is $ 3 $) has more fluctuations and low convergence rate compared to the red curve where the number of subcarriers is $ 2 $. Note that the subcarrier assignment action space is $ 2^{N\times M} $, ($ N $ and $ M $ are the number of subcarriers and users), respectively. Furthermore, Fig. \ref{Loss} presents the loss value function in which the y-axis (loss function) is in the logarithmic scale. 
	This figure shows that the loss function  approximately converges to  zero after  150   episodes.
	The reason behind  the fluctuations in the figure  can be explained by: 1) the logarithmic scale of the figure, and 2)  the impact of the neural network parameters, mainly the batch size and the learning rate, and also the fact that the estimated values for the reward function  may not equal to the actual values.

Below, we discuss the results of the simulations to further analyze and compare the proposed system model with respect to the baselines and investigate the impact of the main network parameters changes in sequence. For the convenience of comparison between the results, the curves with the same color but different marks are for the same parameters, i.e., solid lines are for the proposed algorithm and dash lines are for the baseline.
\subsubsection{Comparing With Baseline 1}
\begin{figure}[t]
	\begin{center}$
		\begin{array}{cc}
		\subfigure[h][	Objective function  versus power budget ]{
			\includegraphics[width=0.5\textwidth]{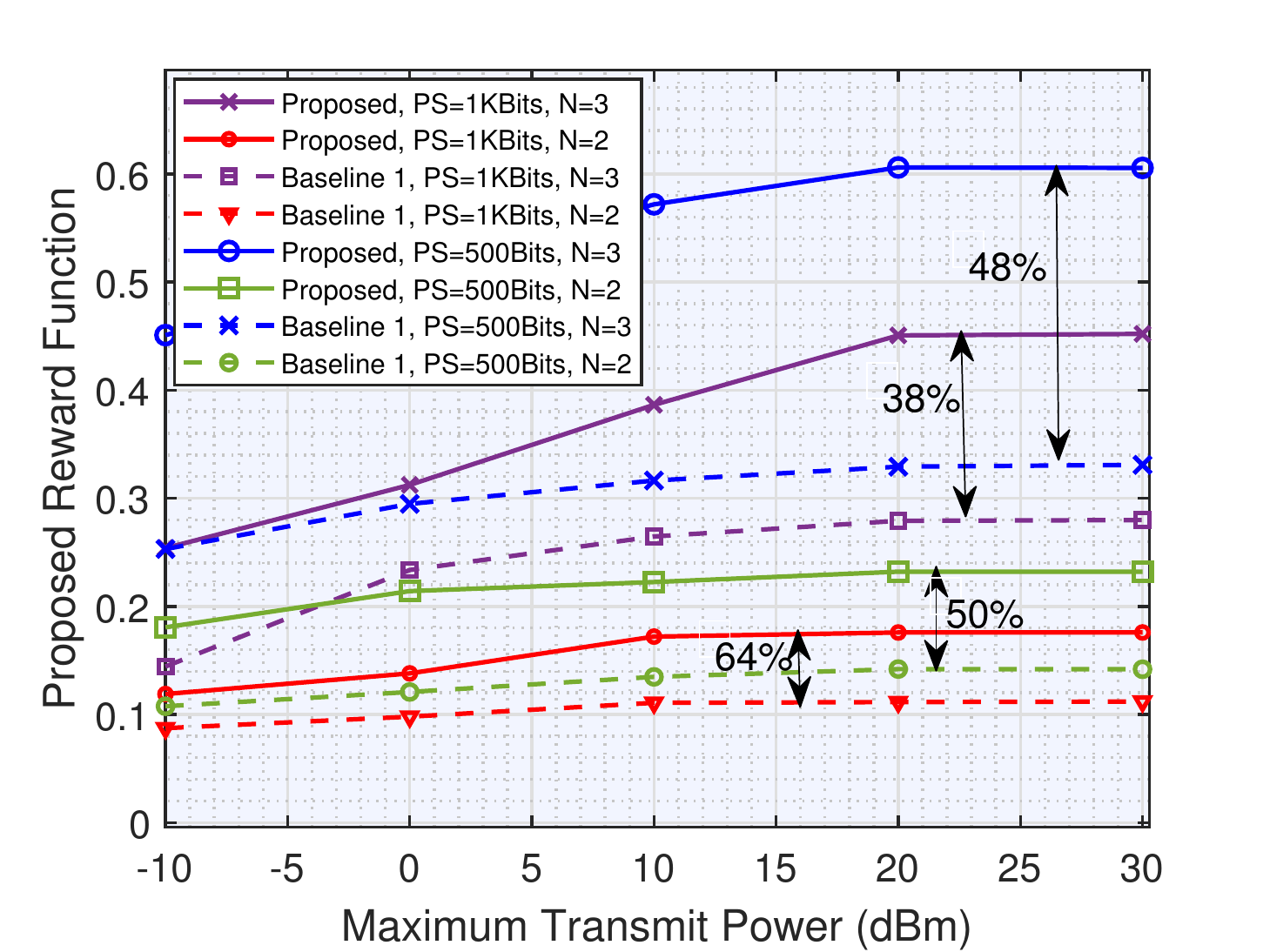}
			\label{Reward_NOMA_OMA_DRL}}
		\subfigure[AAoI versus power budget]
		{
			\includegraphics[width=0.5\textwidth]{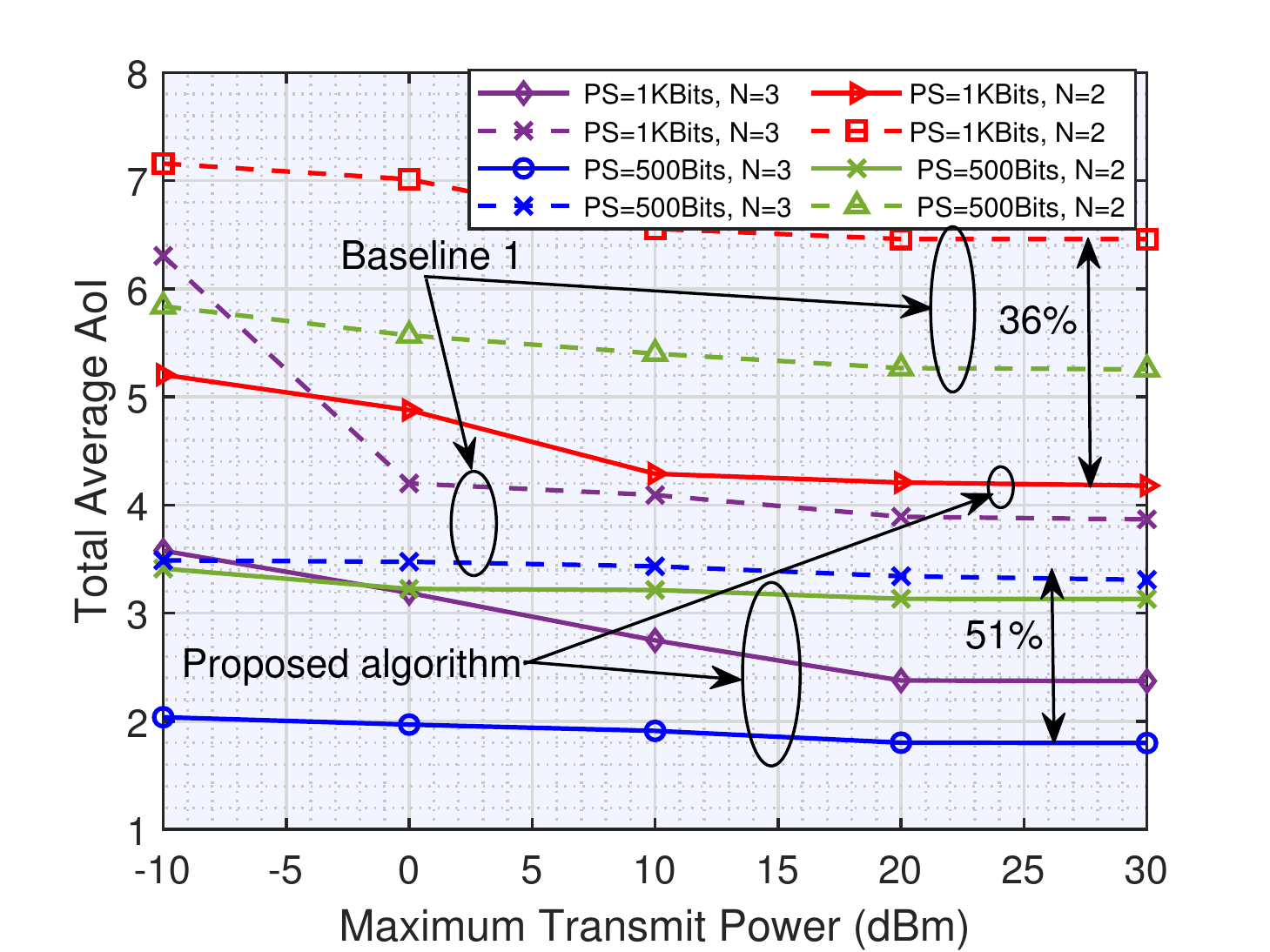}
			\label{AAoI_NOMA_OMA_DRL}}
		%
		\end{array}$
	\end{center}
	\vspace{-1em}
	\caption{Evaluation of objective (ESE over AAoI) and AAoI under variation of the maximum allowable transmit power for each user.
	 }
\end{figure}

\begin{figure}[t]
	\begin{center}$
		\begin{array}{cc}
		\subfigure[h][	Packet dropping versus power budget]{
			\includegraphics[width=0.5\textwidth]{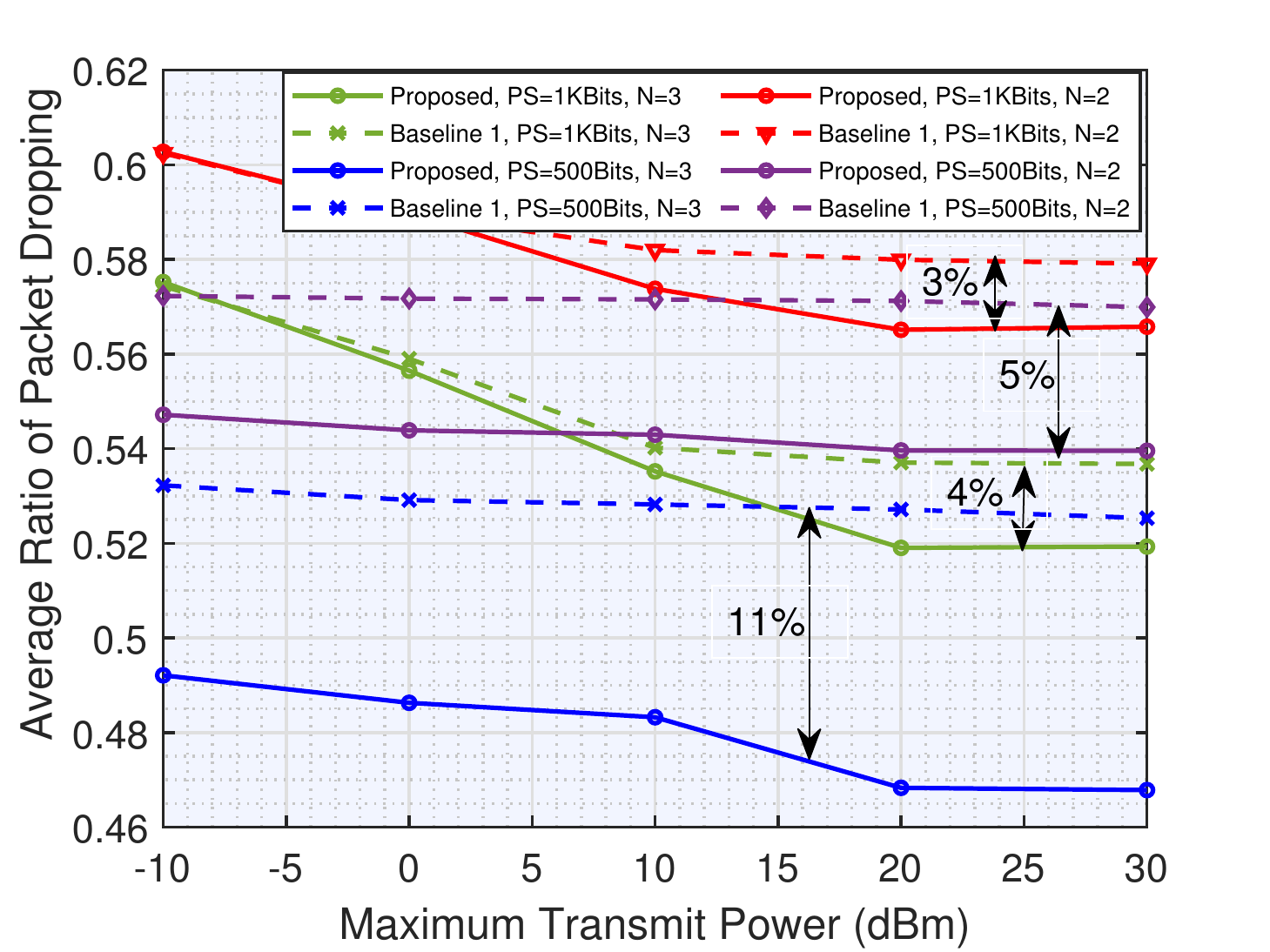}
			\label{Drop_NOMA_OMA_DRL}}
		\subfigure[EE versus power budget]
		{
			\includegraphics[width=0.5\textwidth]{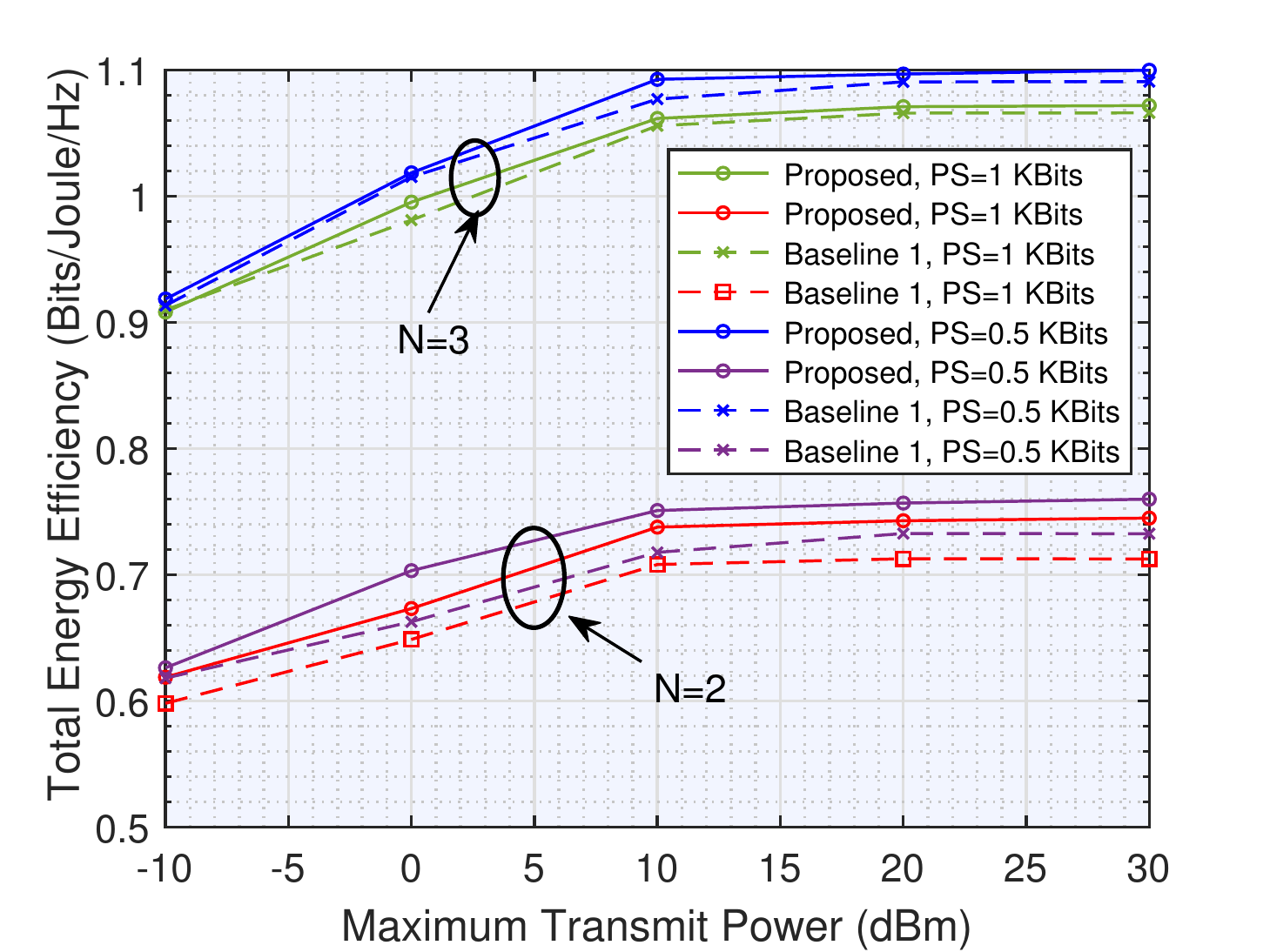}
			\label{EE_NOMA_OMA_DRL}}
		%
		\end{array}$
	\end{center}
	\vspace{-2em}
	\caption{Evaluation of 	ratio of the  packets dropping and EE versus the maximum allowable transmit power per user. 
	}
\end{figure}

The results related to this subsection are in Figs. \ref{Reward_NOMA_OMA_DRL}--\ref{EE_NOMA_OMA_DRL}.  
First, we investigate the proposed objective function, i.e., reward function versus the the maximum allowable transmit power\footnote{For simplicity, we state power budget or power instead.} for each user which is depicted in Fig. \ref{Reward_NOMA_OMA_DRL}. In this figure, we also evaluate the effect of the number of subcarriers and different packet sizes (PS) on the reward function. It is observed that the proposed algorithm  outperforms for all PSs and number of subcarriers. The reward function is increased as the power budget, i.e., $ p_{{m}}^{\max} $, is increased and finally converges to a maximum point. This goes back to the structure of the reward function which is composed of two parts: 1) ESE and 2) AoI. At the beginning, the ESE increases linearly with power budget, until it reaches to a saturated point due to intelligent power control in our algorithm. Also, SE (or throughput) increases which results in a high  number of packet transmissions in the network. As a result, if we have intelligent packet transmission decision regarding to the maximum AoI values, system's total AoI will decrease. 
It is worthwhile to note that network SE and AoI have a direct relationship with each other, not necessarily linear.  We have two main factors 
influencing the SE which are availability of power and spectrum resources. Theoretically, without limitations on the resources, we can achieve a high amount of throughput with minimum total AoI. But in practice, we have  different factors such as regulatory, battery life, human health, cost,  each of which poses numerous limitations on the utilization of the resources. Hence, considering only SE, AoI, or EE as the network performance metric, it is similar to ignoring some restrictions that are critical. But, our objective function is affected by all those performance metrics and also maintains acceptable values for them.

Fig. \ref{AAoI_NOMA_OMA_DRL} presents the AAoI values in terms of the power budget.  It can be seen the superiority of the proposed algorithm in terms of the AAoI and indirect relationship between the AoI and power budget. By increasing the power budget, the throughput of the system is increased. This results in more packet transmissions to the server side which leads to the fresher information packets at the server and reduced levels of AAoI. However, care must be taken since more power budget results in more power consumption which increases the network cost. At the same time, exploiting NOMA enables the capability of spectrum reuse which can increase the network SE. Hence, we have lower AoI as stated before. This is the reason for superiority of the proposed algorithm.
\\\indent The ratio of the packet dropping defined as $ \frac{\text{The number of transmitted packets}}{\text{All generated packets}} $ is  investigated as shown in Fig. \ref{Drop_NOMA_OMA_DRL}. Significant  impact of power budget and NOMA on dropping reduction can be seen from the figure.  Also, it can be implicitly conceived that if packet dropping has not been considered, more power and spectrum was needed to be utilized to send buffered packets, struggling to reduce the AAoI while wasting the network resources to send old information.

The results on the total EE  of the network versus the power budget are drawn in Fig. \ref{EE_NOMA_OMA_DRL}. As seen from the figure, the PS has a negligible impact on the total EE while it is highly affected by the number of subcarriers (utilized bandwidth) and power resources. Note that the ``Hz'' is used due to the normalization of the EE over the subcarrier spacing. 
\subsubsection{Comparing With Baseline 2}
\begin{figure}[t]
	\begin{center}$
		\begin{array}{cc}
		\subfigure[h][	Reward function  versus power budget ]{
			\includegraphics[width=0.5\textwidth]{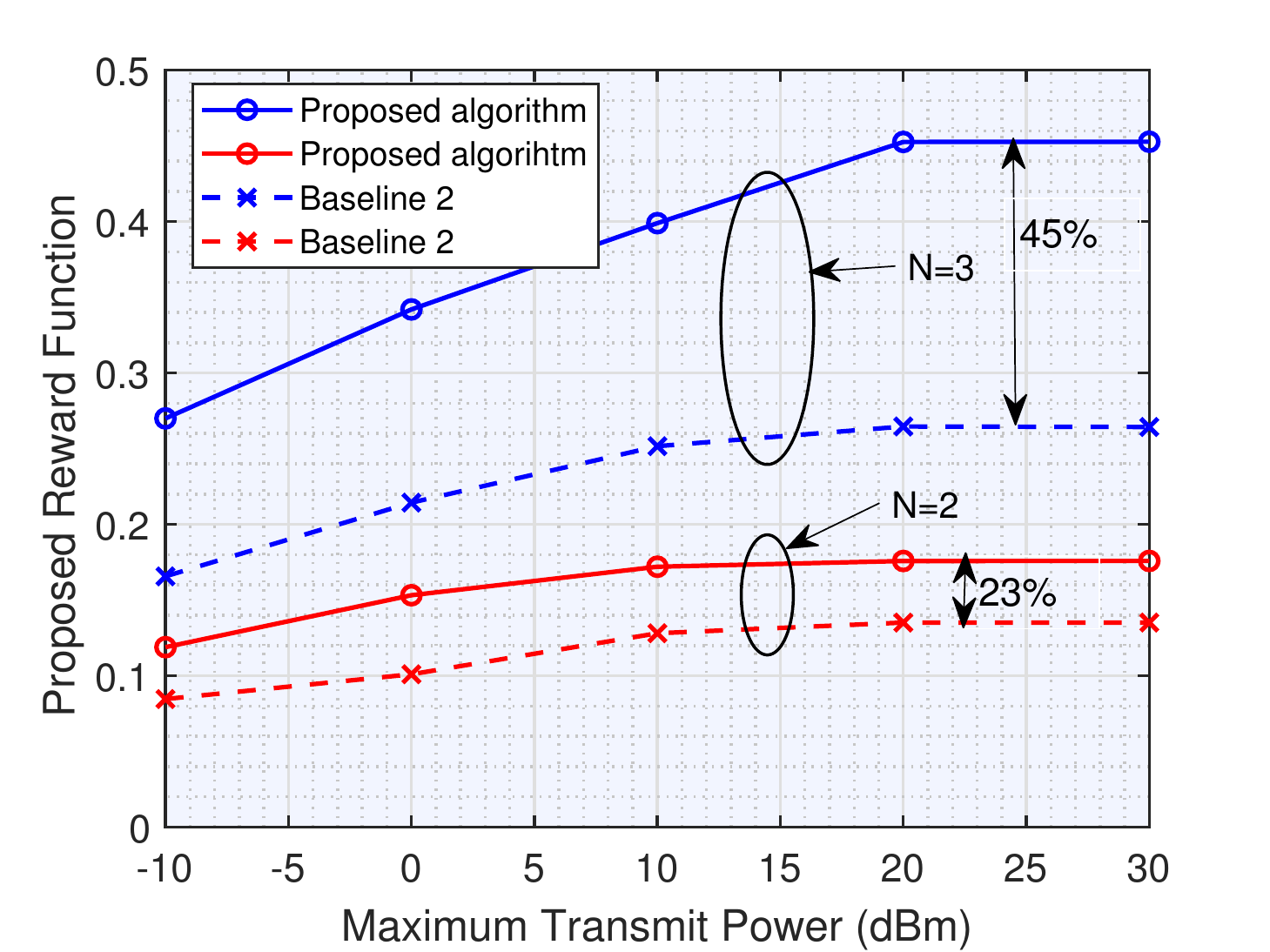}
			\label{Reward_Matching}}
		\subfigure[AAoI versus power budget]
		{
			\includegraphics[width=0.5\textwidth]{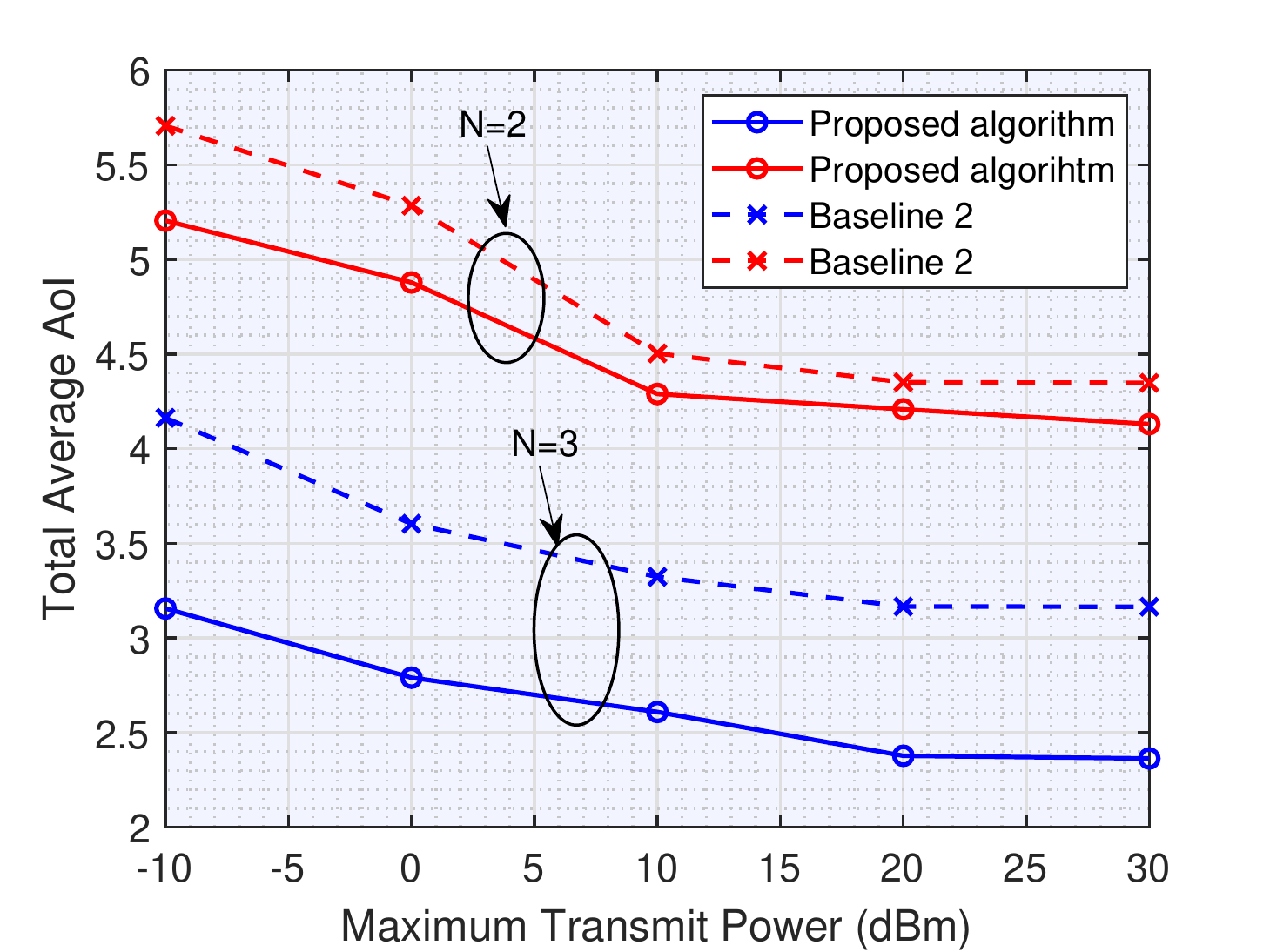}
			\label{AAoI_Matching}}
		%
		\end{array}$
	\end{center}
	\vspace{-1.5em}
	\caption{Evaluation of reward function and AAoI versus to the variation of the maximum allowable transmit power for each user for baseline 2.
	}
\end{figure}

The comparison between the proposed algorithm and the  matching algorithm from the AAoI and reward function perceptive under the variation of power budget is provided in Fig. \ref{AAoI_Matching} and Fig. \ref{Reward_Matching}, respectively. From the figures, we observe that our proposed algorithm outperforms for all the  values of the power budgets and the number of subcarriers. Also, we see that the subcarrier allocation management has a significant impact on the AAoI as well as ESE which are unified in the proposed reward function.   

\subsubsection{Comparing With Baseline 3}
\begin{figure}[!ht] 
	\centering
	\includegraphics[width=.45\textwidth]{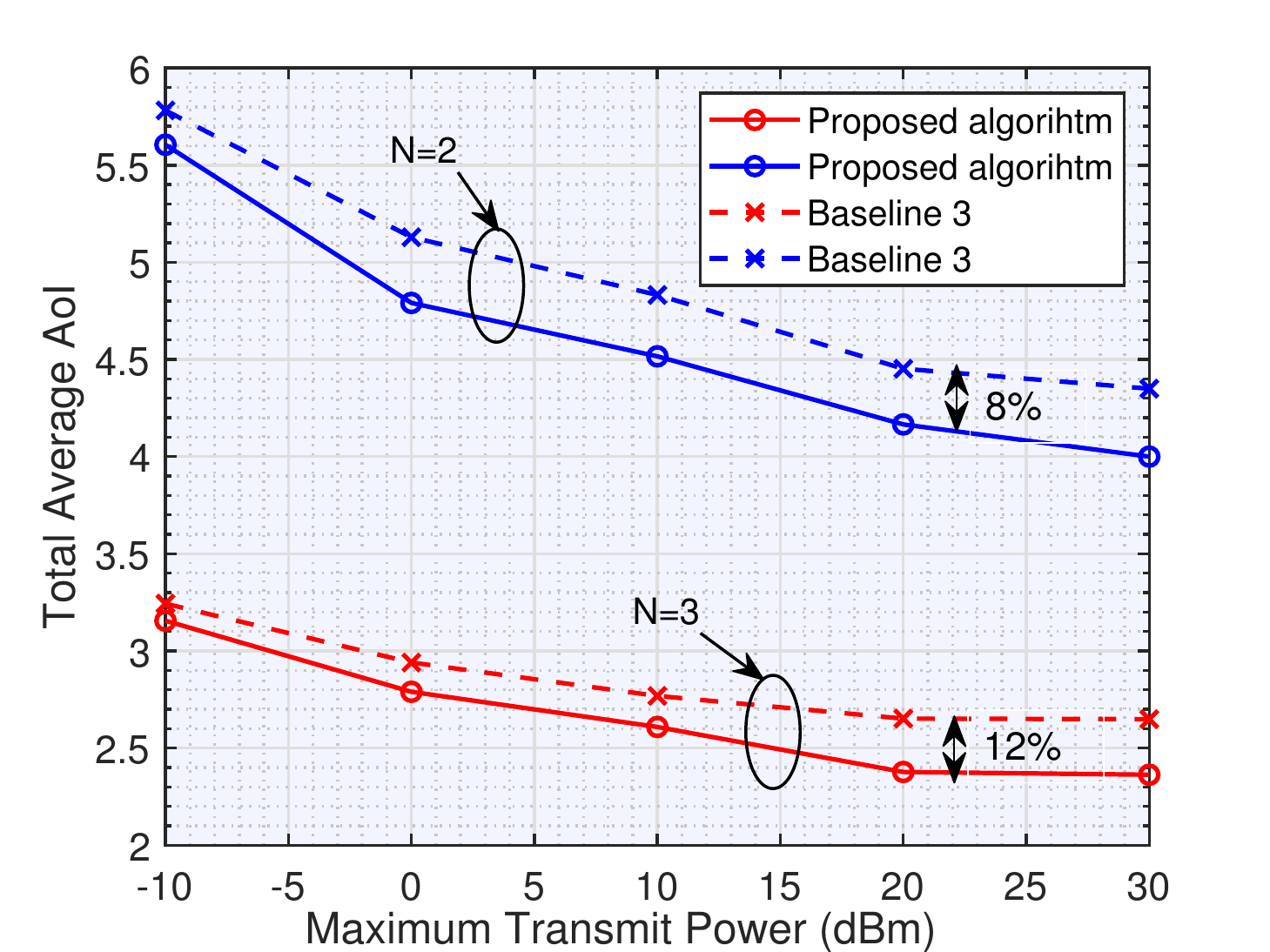}
	\caption{	
		Total AAoI versus the maximum allowable transmit power per user. 
	} 
	\label{AAoI_UPA_NOMA_}
\end{figure}

\begin{figure}[!ht] 
	\centering
	\includegraphics[width=.45\textwidth]{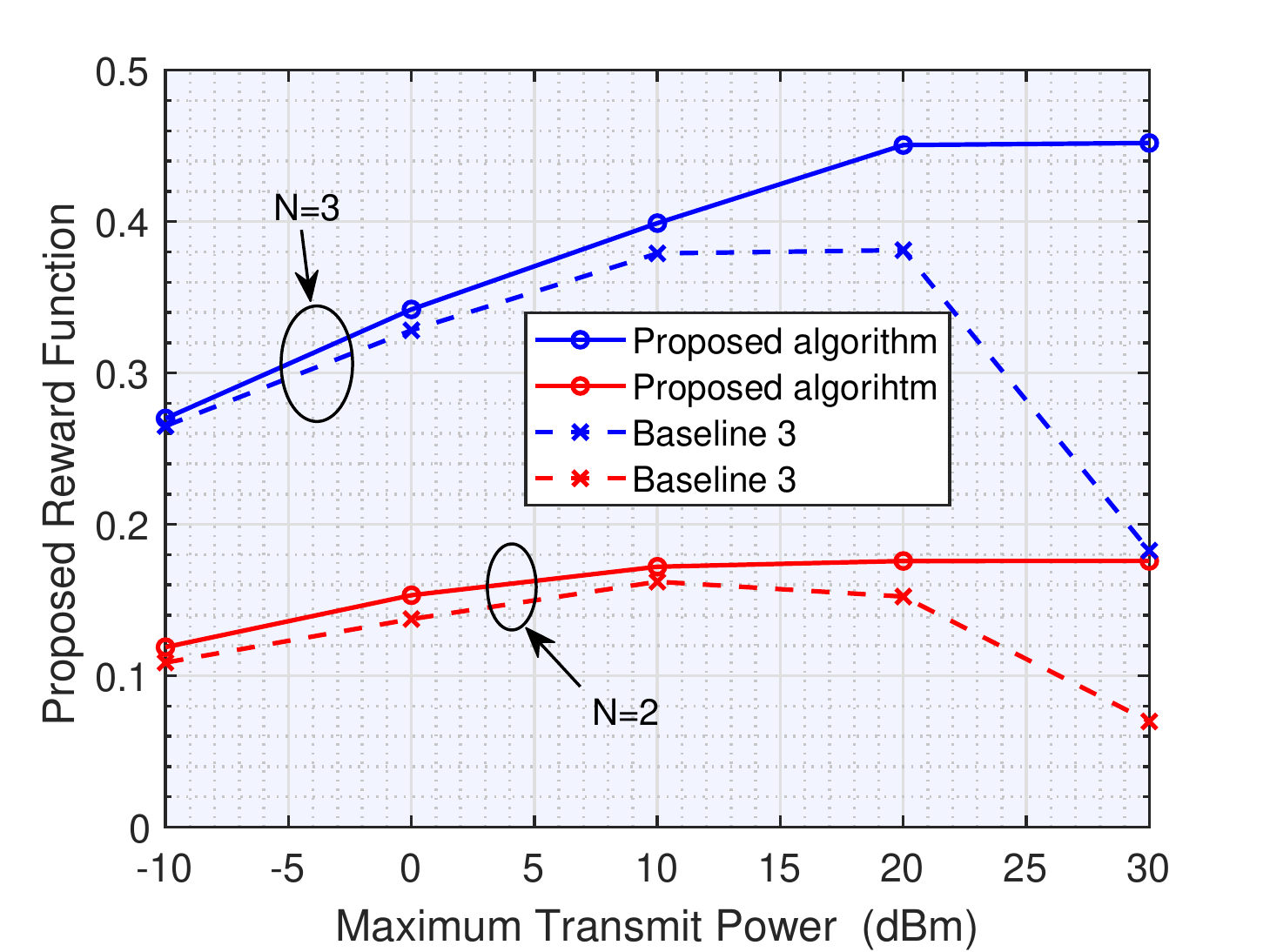}
	\caption{	
		Reward function versus the maximum allowable transmit power for each user. 
	} 
	\label{Reward_UPA_NOMA_}
\end{figure}

\begin{figure}[!ht] 
	\centering
	\includegraphics[width=.45\textwidth]{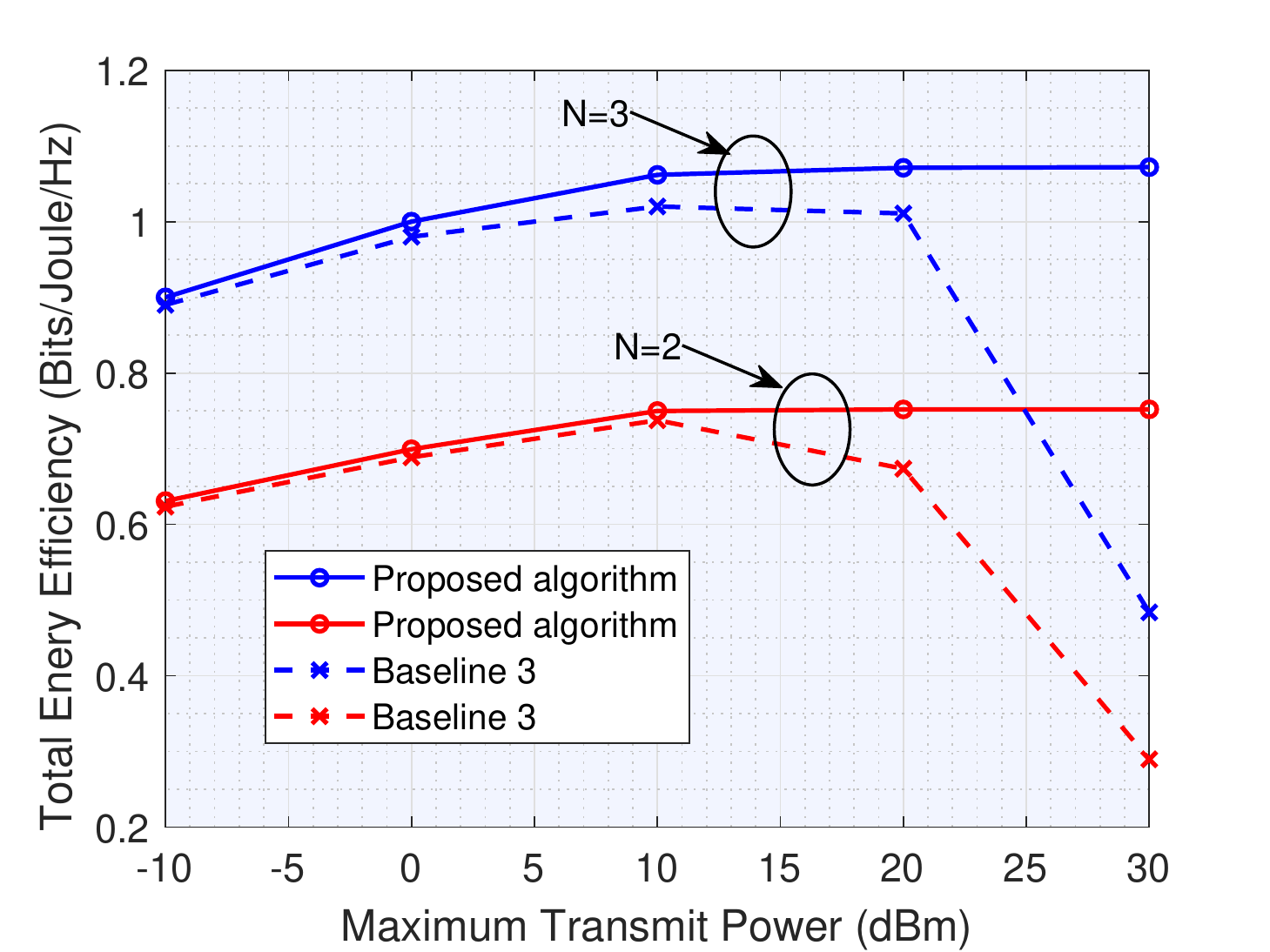}
	\caption{	
		EE versus the maximum allowable transmit power for each user. 
	} 
	\label{EE_UPA_NOMA_}
\end{figure}

First, in Fig. \ref{AAoI_UPA_NOMA_} the total AAoI versus the power budget is plotted. In this figure, the PS is fixed to  1 Kbits and the number of users is 5.  We observe  the considerable impact of the number of subcarriers and superiority of the proposed algorithm. The reason for these are in two-folds: 1)  increasing the number of subcarriers means that we have more bandwidth, (each subcarrier has $ 60 $ KHz bandwidth), hence, the throughput of the network is increased. In such a case, we can transmit more packets which leads to the reduction in AoI level; and 2) via optimizing power allocation, especially for high power budgets, for the fixed parameters, we can also achieve high throughput which results in more packet transmission and decreasing AAoI values. This figure illustrates the important effect of the power allocation algorithm on the AAoI.
\\\indent The reward function versus the power budget and the effect of the number of subcarriers on the reward function are  studied in Fig. \ref{Reward_UPA_NOMA_}. As seen from the figure, the reward function (total ESE over AAoI) is increased as we increment the power budget and finally reaches a maximum value (saturated point) for the proposed algorithm. But this behavior is not observed for the baseline 3 where after a certain power budget value $ P_{m}^{\max}= 20$ dBm,  reward drops and the gap between the proposed algorithm and this baseline becomes larger. 
The reason behind this behavior  is that by increasing power budget the utilization of power increases linearly while the  SE increases by log-function.
 Comparing this figure with  Fig. \ref{AAoI_UPA_NOMA_}, we obtain that this behavior emanates from the EE. In this regard, we evaluate the EE for this case in Fig. \ref{EE_UPA_NOMA_}. As  expected, the ESE drops for power values more than 20 dBm. This is because the increase in power budget leads to more power consumption and hence ESE reduction.  It is worthwhile to note that by increasing the power budget the consumed power is increased  linearly and data rate, especially SE increases with  commensurate to log-function. 
As a result, the proposed reward function can be considered as a unified network performance metric that unveils the impact of various parameters (e.g., PS and available bandwidth) and resource management algorithms on network performance.

\subsubsection{Comparing With Baseline 4}
\begin{figure}[!ht] 
	\centering
	\includegraphics[width=.45\textwidth]{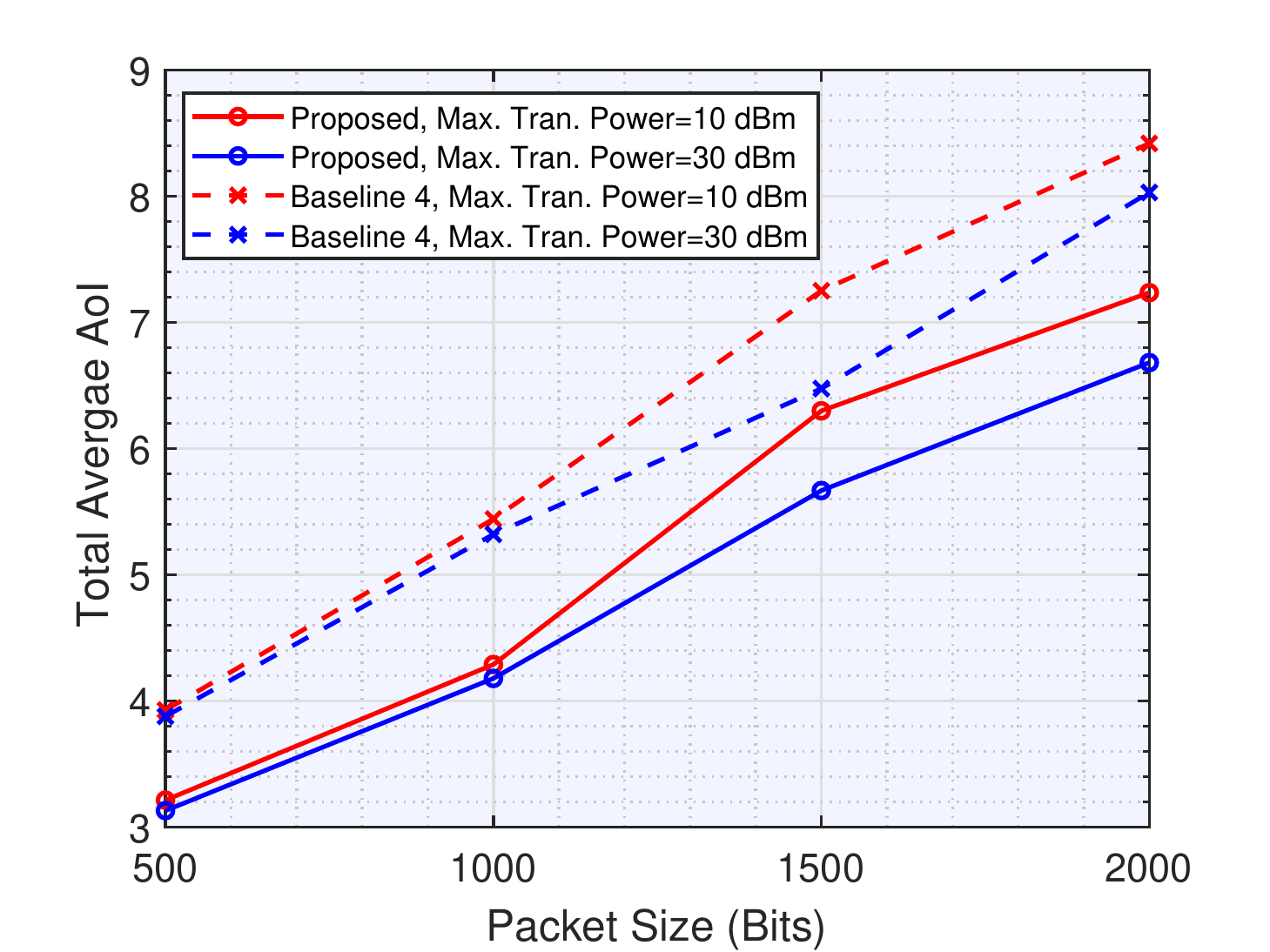}
	\caption{	
		AAoI versus the packet size.
	} 
	\label{AAoI_PS}
\end{figure}

Fig. \ref{AAoI_PS} illustrates the impact of PS and power budget on the total AAoI for the proposed algorithm and baseline 4. From this figure, two key points can be deduced: 1) the significant impact of PS on the AAoI such that increases linearly with PS; and 2) the effectiveness of the proposed algorithm and power budget on AAoI reduction, specifically for larger PSs. The reason behind point ``1'' is that for a given throughput, for short PSs, we can transmit packets. Hence, the generated information is received at the server side more frequently which results in AAoI improvement. This inspires that for future communication networks to support diverse applications with new stringent requirements these it is a necessity to redesign all layers of the communication protocol stack.   

\vspace{-3em}
	\section{Conclusion Remarks}\label{Conclusion}
This paper proposed a  novel unified  performance metric that is defined by ESE over AAoI in a joint radio and computing resource management framework considering multiple types of information for each user, exploiting multi-carrier NOMA and MEC in the uplink communication. In this regard, the CMDP is formulated to optimize dynamically and jointly power, spectrum, and packet transmission decision.  Also a  DRL algorithm under AoI, NOMA, and resource capacity constraints is developed to solve the long-term stochastic optimization problem in an intelligent manner. Simulation results are extensively provided to assess the impact of different parameters, e.g., PS and power budget, and to compare with the state of the art baselines.  The results illustrate that the proposed algorithm outperforms  the baselines and show the significant network performance  in terms of the PS, available bandwidth, and ML-based algorithms. 

	\bibliographystyle{ieeetr}
	
\end{document}